\documentclass{aa}  

\usepackage{graphicx}
\usepackage{txfonts}
\usepackage{lipsum}
\usepackage{amsmath}	
\usepackage{subcaption}
\usepackage{subcaption}         
\usepackage{lscape}             
\usepackage{placeins}           
\usepackage{makecell}
                                
\usepackage{hyperref}

\begin{document}

   \title{Sub 35-ps Single-Photon Stellar Intensity Interferometry of Vega 
   }
    \titlerunning{35-ps Single-Photon SII of Vega}
   \subtitle{}

    \author{Verena G. Leopold\inst{1,3}$\ddagger$\thanks{E-mail: vero.leopold@fau.de},
Sebastian Karl\inst{1}$\ddagger$\thanks{E-mail: seb.karl@fau.de},
Stefan Richter\inst{1,3},
Jean-Pierre Rivet\inst{2},
Yury Prokazov\inst{3},
Evgeny Turbin\inst{3},
Gennady Sintotskiy\inst{3},
Iurii Datii\inst{1},
Jason Aufdenberg\inst{4},
and Joachim von Zanthier\inst{1}}
\institute{Friedrich-Alexander-Universität Erlangen-Nürnberg (FAU), Quantum Optics and Quantum Information, Staudtstr. 1, 91058 Erlangen, Germany
\and Université Côte d'Azur, Observatoire de la Côte d'Azur, CNRS, Laboratoire Lagrange, France \and  
Photonscore GmbH, Magdeburg, Germany  \and Embry-Riddle Aeronautical University, Department of Physical Sciences, Daytona Beach, FL, United States\\
}
 \authorrunning{V. G. Leopold, S. Karl et al.}


   \date{Received \today}

 
  \abstract
   {Stellar intensity interferometry (SII) enables measurements of stellar photospheres by correlating intensity fluctuations, offering a robust alternative to amplitude-based techniques. With the advent of modern single-photon counting detectors, this method has re-emerged as a powerful technique in stellar interferometry exploiting their ultra-high timing resolution.}
    {We aim to determine the stellar diameter of Vega at 405\,nm using spatial intensity correlations measured at the C2PU facility, part of the Observatoire de la Côte d'Azur in southern France.} 
   {We employed an optical and mechanical setup identical to that outlined in \cite{Leopold2025}, involving an ultra-narrow bandpass filter centered at 405\,nm and a polarizing beam splitter. The two campaigns in April and August 2024 used two different detectors, namely hybrid photodetectors (HPDs) with a cross-correlation timing resolution of 22.7\,ps and newly developed PhotonPix detectors with a cross-correlation timing resolution of 34.8\,ps, respectively. The data acquisition was facilitated by a time-to-digital converter from Photonscore that integrated photon arrival times and IRIG-B timestamps for synchronization. Our analysis consisted of calculating auto- and cross-correlation histograms from the recorded data, which were then normalized and corrected for optical path differences.}
  {The temporal auto-correlations measured in April and August 2024 yield a consistent coherence time of 0.21\,ps. This confirms the effectiveness of the PhotonPix detectors despite a slight degradation in timing resolution relative to the HPDs. Combining the data from the two measurement campaigns yields a uniform disk diameter of Vega of $2.97 \pm 0.23$\,mas. This value agrees well with the uniform disk diameter of 3.047\,mas predicted by the stellar atmosphere model at our observation wavelength, validating both our measurement technique and the new PhotonPix detectors for precision SII measurements at sub-35\,ps timing resolution.}
 {The integration of advanced single-photon detectors like the PhotonPix used for the first time in the current experiment into existing Imaging Atmospheric Cherenkov Telescope SII instruments promises substantial enhancements in signal-to-noise ratio and observational capabilities for SII, particularly with telescopes like the Cherenkov Telescope Array Large Size Telescopes. Although challenges such as night sky background saturation and precise positioning and guiding remain, SII observations at apparent magnitudes up to 7.8 are within reach at these telescopes.}


   \maketitle
\nolinenumbers 
    \def\thefootnote{$\ddagger$}\footnotetext{These authors contributed equally to this work}\def\thefootnote{\arabic{footnote}}

\section{Introduction}
Stellar intensity interferometry (SII) was first demonstrated for visible starlight by \cite{HanburyBrown1956}. In this seminal experiment, Hanbury Brown \& Twiss (HBT) measured the diameter of the brightest star in the night sky, Sirius ($\alpha$ CMa). Compared to the amplitude interferometry of \cite{Michelson1921}, intensity interferometry avoids the technical challenge of coherently superposing light from two different telescopes. Instead, light intensities are detected locally at the respective telescopes and the corresponding electrical or digital signals are centrally correlated. The simplicity of SII allowed \cite{Brown1967} to construct the first large-scale stellar interferometer, the Narrabri Stellar Intensity Interferometer (NSII). At the NSII, \cite{HanburyBrown1974a} measured the diameters of 32 stars. However, around the same time, \cite{Labeyrie1975} detected the first amplitude interferometry fringes of Vega with two separate telescopes. Due to its higher sensitivity per detected photon, amplitude interferometry subsequently replaced SII and is currently the workhorse of stellar interferometry. Instruments like GRAVITY at the VLTI reach magnitudes of $>15$ \citep{GRAVITY+Collaboration2022}, enabling interferometry of extra-galactic sources, while the MIRC-X beam combiner at the CHARA array combines light from six telescopes simultaneously for imaging \citep{Anugu2020}. 

Recently, SII has received renewed interest due to significant advances in photon detection technology. In this context, \cite{Dravins2013} were among the first to recognize the potential of SII when combined with modern arrays of telescopes with large diameters like the Cherenkov Telescope Array (CTA). Here one noteworthy potential contribution is the more dense sampling of stellar visibilities if a large number of telescopes is used, providing camera-like "imaging" at high resolution. The first SII implementation between two telescopes since the NSII was performed by \cite{Guerin2018} at the twin $1\,$m optical telescopes of the C2PU facility, part of the Observatoire de la Côte d'Azur. SII implementations at major Imaging Atmospheric Cherenkov Telescope (IACT) arrays soon followed at MAGIC \citep{Acciari2019a}, VERITAS \citep{Abeysekara2020}, and H.E.S.S. \citep{Zmija2023}. These SII instruments reach the level of maturity that is necessary to perform stellar diameter surveys \citep{Kieda2021a, Abe2024} and multicolor observations \citep{Vogel2024}. All SII implementations at IACTs to date operate at blue wavelengths and use conventional photodiodes, which return a photocurrent for light detection. 

The newly constructed CTA Large Size Telescopes (LSTs) are special within the IACT category: their active mirror control allows for them to be segmented into sections with anisochronicities $\le100\,$ps \citep{Cortina2025}. Their photon timing dispersion thus stays smaller than that of their current electronic signal-acquisition pipeline by a factor of $\ge 80$ \citep{Abe2024}. This means that the SII instruments currently installed at these facilities fall short of their maximum possible signal-to-noise ratio by a factor of $\ge 5$ \citep{HanburyBrown1974, Dalal2024, Guerin2025}. Single-photon detectors can overcome this bottleneck, while also being less sensitive to electronic noise. In practice, however, the large focal spot of large IACTs ($\approx 30\,$mm at the focal plane) and the prohibitively large detected photon rate ($>100\,$MHz when observing a mag. 7 star \cite{Cortina2025}) have so far prevented the adoption of single-photon detectors at IACTs.

We have recently developed and tested two components: (1) a commercial large-area single-photon detector sustaining detection rates up to $100\,$MHz at visible wavelengths, together with its digitization electronics (PhotonPix and LINTag, respectively, both from Photonscore GmbH), see \cite{Karl2025_pub}; and (2) an SII instrument optimized for large-area single-photon detectors \citep{Leopold2025}. In this paper, we combine these two achievements to measure the diameter of Vega ($\alpha$ Lyr.) with 82\,ps timing resolution, and evaluate the feasibility of using single-photon detectors for SII at IACTs with low photon timing dispersion.
 
\section{Spatial Intensity Interferometry}
\FloatBarrier
The theory of SII can be described in terms of classical thermal fields for starlight, see \cite{HanburyBrown1974} for an in-depth treatment. SII measures the second-order intensity correlation function 
\begin{equation}
        g^{(2)}(r, \tau) = \frac{\left< I(r_0, t_0) I(r_0 + r, t_0 + \tau)\right>}{\left< I(r_0, t_0) \right> \left< I(r_0 + r, t_0 + \tau)\right>},
        \label{eq:1}
\end{equation}
where $I$ is the instantaneous intensity at a given spacetime point, $r$ is the projected baseline between two telescopes positioned at $r_0$ and $r_0 +r$, respectively, $\tau$ is a temporal delay, and $\left< \right>$ denotes time-averaging. The second-order correlation function can be interpreted as the conditional probability of detecting a photon at position $r_0 + r$ and time $t_0 + \tau$ after having previously detected a photon at position $r_0$ and time $t_0$ \citep{Fox2006}. We use this to measure $g^{(2)}(r, \tau)$ by digitizing and correlating single-photon arrival time streams at two distant telescopes.

\begin{figure}
    \centering
    \includegraphics[width=\hsize]{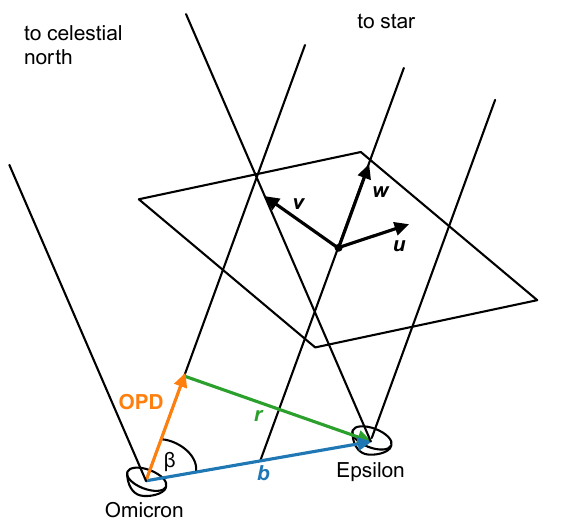}
    \caption{Coordinate systems used in stellar interferometry following the conventions in \cite{Thompson2017}, drawn for the C2PU telescopes (see Sec.~\ref{sec:telescope}). The optical path delay ($\vec{\mathrm{OPD}}$) is shown in orange for a given absolute baseline $\vec{b}$ shown in blue, and the projected baseline $\vec{r}$ shown in green. The projected baseline $\vec{r}$ lies inside the $uv$-plane, where $\vec{v}$ points towards celestial north and the normal vector $\vec{w}$ points towards the star.}
    \label{fig:hbt_coords}
\end{figure}

Since the two telescopes are placed at fixed positions separated by a fixed baseline $\vec{b}$, the star's motion across the sky changes the projected baseline $r = \left| \vec{r} \right|$ \citep{Labeyrie2006}. We show the typical geometry and coordinate systems for SII, adapted to the telescope site, in Fig.~\ref{fig:hbt_coords}. Photons emitted by the star at the same time can be delayed at one telescope (Omicron in Fig.~\ref{fig:hbt_coords}) with respect to the other by a distance called the optical path delay OPD $= \left| \vec{\mathrm{OPD}} \right|$. During measurement acquisition, we need to compensate for the change in OPD by shifting the correlation histograms accordingly. Calculating the OPD and projected baseline is straightforward if we know the angle $\beta$ between $\vec{b}$ and the unit vector $\vec{w}$ that points towards the star, since the OPD, $\vec{b}$, and $\vec{r}$ form a right triangle. The projected baseline $\vec{r}$ always lies inside the plane with normal vector $\vec{w}$, called the $uv$-plane. Inside this plane, we choose basis vectors $\vec{v}$ (that points towards celestial north) and $\vec{u}$ to make up a complete right-handed coordinate system. In this paper, we thus follow the definition of the $u,\, v, \, w$ coordinate system by \cite{Thompson2017}.

To connect $g^{(2)}(r, \tau)$ with source properties, we rewrite it in terms of the spatial and temporal first-order correlation functions, denoted $g^{(1)}(r)$ and $g^{(1)}(\tau)$ respectively, \citep{Mandel1995}:
\begin{equation}
    g^{(2)}(r, \tau) = 1 + \left| g^{(1)}(r) \right|^2 \left| g^{(1)}(\tau) \right|^2.
    \label{eq:2}
\end{equation}
Eq.\,(\ref{eq:2}) relies on the Siegert relation \citep{Siegert1943}, which is valid for thermal light. Both $g^{(1)}(r)$ and $g^{(1)}(\tau)$ are related to source properties via Fourier transform relationships, called the van Cittert-Zernike theorem and the Wiener-Khintchine theorem \citep{Mandel1995}. The van Cittert-Zernike theorem states that $g^{(1)}(r)$ can be derived from the Fourier transform of the source intensity distribution, and the Wiener-Khintchine theorem expresses that $g^{(1)}(\tau)$ can be calculated by the Fourier transform of the normalized source spectrum. Using the simplest possible model for a star, a circular disk of uniform intensity with angular diameter $\delta$, we obtain 
\begin{equation}
    g^{(1)}(r) = 2 \frac{J_1(\pi r \delta / \lambda)}{\pi r \delta / \lambda},
\label{eq:3}
\end{equation}
where $J_1$ is the Bessel function of first kind and first order, and $\lambda$ is the central observation wavelength. We use this result to construct the uniform disk (UD) model fit function that we employ for fitting $\left| g^{(1)}(r) \right|^2$. More specifically, suppose we have acquired multiple measurements of $ g^{(2)}(r_n, \tau)$ at fixed star positions on the night sky, which correspond to certain projected baselines $r_n$. We define the coherence time $\tau_c$ as the area under the peak of the second-order temporal auto-correlation function
\begin{equation}
   \tau_c = \int \left[g^{(2)}(r= 0, \tau) - 1\right] \, d\tau = \int \left| g^{(1)}(\tau) \right|^2 \, d \tau.
\end{equation}
Thus, temporal integration and rearranging of Eq.\,(\ref{eq:2}) yields
\begin{equation}
    \tau_c \; \left| g^{(1)}(r_n) \right|^2  =  \int \left[g^{(2)}(r_n, \tau) - 1\right] \, d\tau. 
\end{equation}
As a notational shorthand, we call $\tau_c \; \left| g^{(1)}(r_n) \right|^2 \coloneqq \tau_c \, V^2(r_n)$ the visibility curve, where $V$ is the interferometric visibility. To evaluate stellar diameters using the UD model, we thus use a fitting function $f$  
\begin{equation}
    f(r, \tau_c, \delta) = \tau_c \, \left| \frac{2 J_1(\pi r \delta / \lambda)}{\pi r \delta / \lambda} \right|^2,
    \label{eq:fit_func}
\end{equation}
with free parameters $\tau_c$ and $\delta$ to the area under the curve of the second-order temporal correlation functions recorded at different baselines $r_n$.

Without considering excess noise, the signal-to-noise ratio (S/N) of SII, derived by \cite{HanburyBrown1974} for the case of correlating classical photocurrents, has been extended to the single-photon-counting regime by \cite{Dalal2024} and \cite{Guerin2025}:
\begin{align}
        \mathrm{S/N} &=  A \, \epsilon_\mathrm{opt} \epsilon_\mathrm{det} \, n \, \sigma \, \frac{1}{1 + a} \, \left| g^{(1)}(r) \right|^2 \, \sqrt{\frac{b_\nu t_{\mathrm{meas}}}{2}} \notag \\ &=  A \, \epsilon_\mathrm{opt} \epsilon_\mathrm{det} \, n \, \sigma \, \frac{1}{1 + a} \, \left| g^{(1)}(r) \right|^2 \, \sqrt{\frac{t_{\mathrm{meas}}}{\kappa \, \Delta t_\mathrm{RMS}}}  .
    \label{eq:10}
\end{align}
Here $A$ is the telescope mirror area, $\epsilon_\mathrm{opt}$ is the product of all optical efficiencies of the detection setup, $\epsilon_\mathrm{det}$ is the quantum efficiency of the detectors, $n$ is the incident photon flux, $\sigma$ is a normalized spectral density factor,  $a$ is the fraction of stray light compared to the signal intensity, $b_\nu$ is the frequency bandwidth of the photocurrent detection chain, $\Delta t_\mathrm{RMS}$ is the RMS width of the corresponding temporal resolution of the detection setup, and $t_\mathrm{meas}$ is the measurement time. Going from the 3dB frequency bandwidth to the RMS timing resolution of a detection setup using single-photon counters requires multiplication with a correction factor $\kappa = 4 \sqrt{\pi} \approx 7.09$. All temporal resolutions in this paper are given in RMS units. Note that the S/N as defined in Eq.\,~(\ref{eq:10}) does not depend on the spectral or polarization filtering of the incident light \cite{HanburyBrown1974}.

Even though the S/N of SII is independent of optical filtering, practical SII implementations employ it to reduce the photon load on the detectors while at the same time increasing the coherence time. Yet, this filtering, which is performed at each telescope individually, can introduce systematic errors in the measured coherence time if it is not identical. To quantify this systematic coherence loss we introduce coherence factors $\eta$ that allow us to rewrite Eq.\,~(\ref{eq:2}) to 
\begin{equation}
    g^{(2)}(r, \tau) = 1 + \eta_\mathrm{spec} \, \eta_\mathrm{pol} \, \left| g^{(1)}(r) \right|^2 \left| g^{(1)}(\tau) \right|^2.
    \label{eq:7}
\end{equation}

For the polarization coherence factor $\eta_\mathrm{pol}$ we write  the two polarizations as two Jones vectors where one is rotated with respect to the other by an angle $\theta$:
\begin{equation}
    \eta_\mathrm{pol} = \cos^2(\theta).
        \label{eq:8}
\end{equation}
To calculate the spectral overlap coherence factor $\eta_\mathrm{spec}$ we use the ratio of $\sigma$ as defined in \cite{HanburyBrown1974} under the assumption that the spectral response of the detector does not vary over the narrow spectral detection range:
\begin{equation}
    \eta_\mathrm{spec} = \frac{\sigma_\mathrm{offset}}{\sigma_\mathrm{ideal}}
\end{equation}
Here $\sigma_\mathrm{offset}$ is calculated using the two slightly different filter transmission spectra and $\sigma_\mathrm{ideal}$ is the average of the transmissions assuming that the filter at telescope 1 overlaps perfectly with the filter at telescope 2. For identical filtering at both telescopes, both coherence factors are equal to 1.

\section{Methods}
\subsection{Telescopes}\label{sec:telescope}
Our measurements were carried out at the C2PU facility (UAI code R87), part of the Observatoire de la Côte d'Azur at the Calern plateau in southern France. C2PU is located at $43^\circ 45' 13.2''$\, North	
and $06^\circ 55' 22.7''$ East at an altitude of $1274\,$m. Placed roughly on the east-west line, C2PU offers twin telescopes with a Cassegrain secondary focus at a separation of approximately $15\,$m. The telescope in the west dome is called Omicron, and the one in the east dome, Epsilon. The telescopes use an equatorial yoke mount and have a primary mirror diameter of $1.04\,$m, a focal length of $13\,$m, and a focal ratio of $f/12.5$.

The pointing coverage of both telescopes is similar, but Epsilon's tighter constraints define the observation times for both telescopes. The mounting flange of Epsilon is rotated by $10^\circ$ with respect to the mounting flange of Omicron. In addition, Epsilon's pupil is nearly 24\% smaller than that of Omicron, due to a primary mirror baffle. Moreover, the focus of the Epsilon telescope is approximately $70\,$mm shorter compared to that of the Omicron telescope. The latter constraint can be compensated for by moving the secondary mirror, which shifts the focus appropriately. The absolute positions of the Epsilon and Omicron telescopes within the epoch 2024.72 reference frame were measured in September 2024 in collaboration with the Institut Géographique National. This measurement used the total-station technique and measured the coordinates of both telescopes against a common set of reference pillars. It fixes the 3D baseline vector of the C2PU telescopes to an accuracy of $\Delta b\le 0.24$\,mm for each vector component. This introduces a potential systematic timing error of $\le \sqrt 3 \Delta b/c = 1.4$\,ps in the OPD correction, which is negligible when compared to the timing resolution of our detectors. 

\subsection{Experimental setup}

\begin{figure*}
    \centering
    \includegraphics[trim={4.5cm 1cm 3.4cm 5cm},clip,width=\linewidth]{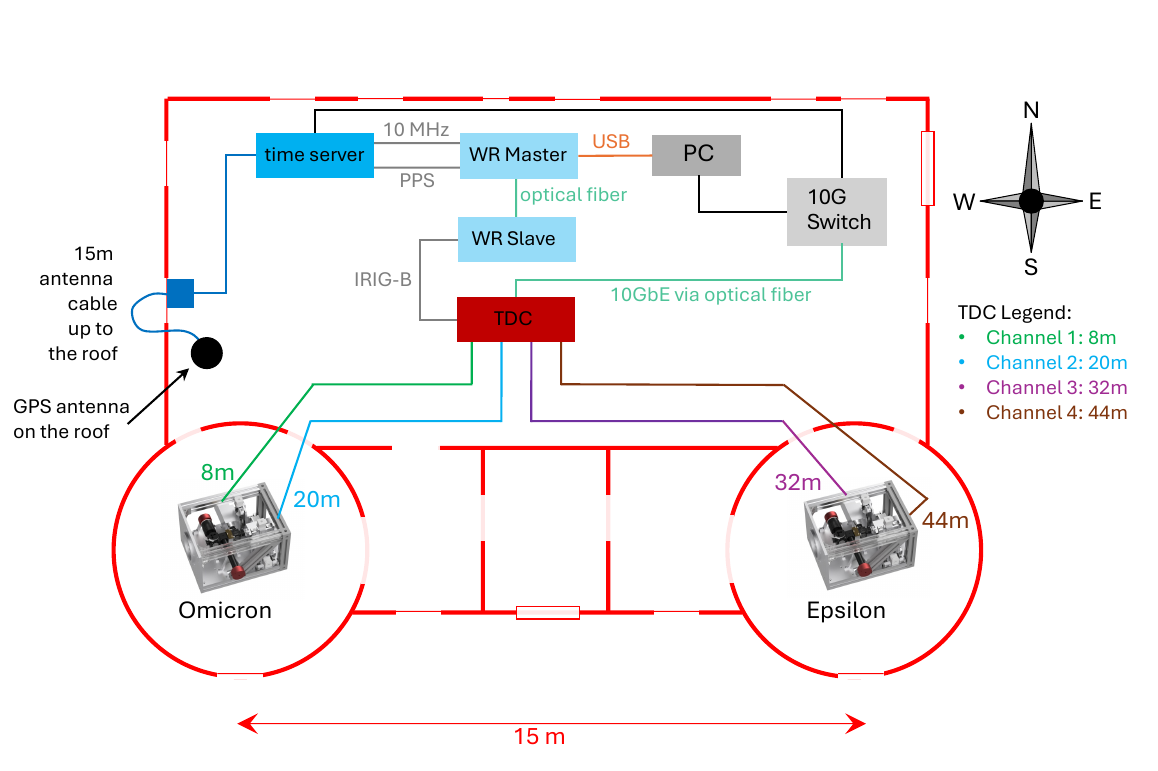}
    \caption{Topology of the measurement setup during the two measurement campaigns in April and August 2024. Our optical setups were mounted to the telescopes in the Omicron and Epsilon domes. The remaining equipment was located in the focal room behind the Omicron dome, but in this sketch it has been expanded to the full extent of the C2PU building for better readability (for details see text).}
    \label{fig:topology}
\end{figure*}

We used the same optical and mechanical setup described in detail in \cite{Leopold2025}. This setup first collimates the starlight, and then filters it using an ultra-narrow bandpass from Alluxa with a 2\,nm full width at half maximum (FWHM) centered at 405\,nm and a polarizing beam splitter. At each telescope, the light is sent to two single-photon detectors via a 50:50 beam splitter to enable measurements of the auto-correlation function at both telescopes. Lenses of appropriate focal length are placed in front of each single-photon detector to converge the collimated and filtered starlight to its active area. To operate two telescopes, we doubled the setup, accounting for the $10^\circ$ difference in mounting on Epsilon to observe the same polarization state at both telescopes. We estimate the maximal angle between the polarizers at the two telescopes to be $\theta \le 1^\circ$ due to tolerances in the flange plate, which leads to a negligible polarization coherence loss of $\eta_\mathrm{pol} \ge 0.9997$. Using high resolution filter spectra obtained from the filter manufacturer, we calculate a spectral coherence factor of $\eta_\mathrm{spec} = 0.996$, which we use in our data evaluation, as detailed below. The loss in coherence is caused by a $0.08$\,nm shift in the filter's central wavelength. In laboratory tests we measured a maximum filter tilt of $\le 0.5^\circ$ in its mount and a beam collimation of $\le 0.05^\circ$ divergence in our setup. Using the theory presented in \cite{Macleod2010} we calculate a central wavelength blue shift of $\le 0.004$\,nm and a filter transmission broadening of $\le0.003$\,nm. These effects can thus be neglected in the calculation of our spectral coherence factor.

We observed Vega using two types of single-photon detectors: In April 2024, we used hybrid photodetectors (HPDs) manufactured by \cite{BHG2017} with an active area of 6\,mm diameter together with custom constant-fraction discriminators (CFDs) for signal discrimination. In \cite{Leopold2025} we measured the quantum efficiency of the HPDs to be 22.7\% at $405$\,nm. We used lenses with a focal length of 40\,mm directly attached to the C-mount threads in the HPDs to converge the light beam. In addition, we observed Vega in August 2024 using the newly developed PhotonPix detectors \cite{Karl2025_pub} to validate their use for SII. These detectors allow for a significantly higher maximum photon count rate (100\,MHz) and a higher quantum efficiency of $\approx 35\%$ at $405$\,nm than the HPDs, and could thus be used for SII at IACTs. The PhotonPix detectors already provide a discriminated output, so no external CFD is needed. Due to the larger active area of 8\,mm diameter, lenses with a focal length of 45\,mm were used to converge the beam. Both detector types have time resolutions of $\le 22$\,ps at detection rates of around 4\,MHz for the individual detectors.

The other parts of the measurement setup were unchanged between our observations of Vega in April and August 2024. We show the scheme of our complete SII instrumentation as installed at C2PU in Fig.~\ref{fig:topology}. All four detectors were connected to the time-to-digital converter (TDC), the LINTag manufactured by \cite{PhotonscoreGmbH2024}. Cables with a mutual delay of at least 12\,m (49.2\,ns) were used to shift the measured correlations out of the crosstalk region of the TDC. Due to attenuation in the cables, different thresholds were set at each TDC channel. The detectors at the Omicron telescope were connected to the TDC on channels 1 and 2, and the detectors at the Epsilon telescope to the TDC channels 3 and 4. The TDC then sends the data to the evaluation PC via a central 10G switch. 

Due to the high timing resolution of the detectors, the OPD correction needs to be accurate to the ps level. We accomplished this by using submillimeter-accurate telescope coordinates and external synchronization of the TDC to UTC time. To this end, we used two White Rabbit (WR; \cite{Lipinski2011}) nodes, one configured as master and one configured as slave. The WR nodes were manufactured by \cite{SED2023}. Each second, the WR slave sends an IRIG-B time code that references the WR internal time to the TDC, where the time code's rising and falling edges are digitized. The WR slave is locked to the WR master via a bidirectional optical fiber link. The WR master clock is disciplined by the clock in an NTP time server (Bürk Mobatime DTS 4160) via its 10\,MHz and pulse-per-second (PPS) coaxial output. The clock in the time server is locked to UTC via a GPS receiver placed on the roof of the observatory building. Unfortunately, the WR master node could not be used as a boundary clock, so we set the internal time of the WR master from NTP time supplied to the PC by the time server via the USB interface, giving us an absolute synchronization accuracy of 1\,s to UTC. Since the OPD of Vega varied by a maximum of $2.8\, \mathrm{ps} / \mathrm{s}$, this error was acceptable, but could be improved by using a WR switch as the master, which directly supports NTP/PTP input. More importantly, the time-code distribution via WR allows extensions of our measurement setup up to a baseline of several km while keeping ps-level synchronization by using a second WR slave node connected to a second TDC. For this application, the TDC oscillator has to be disciplined by the 10\,MHz clock output of the WR slave node, which is supported by the LINTag.

\subsection{Data evaluation}
The data that were recorded to disk on the PC contained photon arrival times for all four detectors and timestamps for the rising and falling edges of the IRIG-B. Data acquisition was chunked into 1-minute intervals to provide manageable time stream file sizes of $<2$\,GiB when observing Vega. We used a custom correlator software written in C\# to decode the IRIG-B timestamps and separate the photon arrival time streams into $1$\,s chunks marked by the respective IRIG-B timestamp. For each $1$\,s chunk, we calculated six different $g^{(2)}(\tau)$ histograms and saved them to disk: the two auto-correlations of detectors mounted at the same telescope (channel pairs: $1 \leftrightarrow 2$, $3 \leftrightarrow 4$) and the four cross-correlations of detectors mounted at different telescopes (channel pairs: $1 \leftrightarrow 3$, $1 \leftrightarrow 4$, $2 \leftrightarrow 3$, $2 \leftrightarrow 4$). These histograms were calculated with a width of 250\,ns and a bin size of 1\,ps to allow for easy histogram shifting during OPD correction. This also requires the histograms to be flat for time differences $\tau$ larger than the dead time of the single-photon detector. This was accomplished by computing them using the multi-start multi-stop technique. This semi-live correlation pipeline can handle total incident photon rates of $> 400$\,MHz and can delete the time-stream file after correlation to save disk space.

When analyzing the correlation histograms, we first normalized both the auto- and cross-correlations with post-recorded normalization histograms measured in our laboratory using white light. For the OPD correction of the cross-correlations, we used the Python package Skyfield developed by \cite{Rhodes2019} to obtain high-precision on-sky positions of the observed star for each 1\,s chunk. For each of these chunks, we computed the angle $\beta$ and calculated the OPD, the projected baseline, and its $u$ and $v$ components, and shifted the four cross-correlations according to the calculated OPD. After processing all histograms of the current observation this way, we added those recorded for each of the six channel pairs and rebinned them to 10\,ps. We call the results of this pipeline the full auto- and cross-correlations. 

Instead of adding up all cross-correlations, we could also choose to add them up in sections based on the hour angle of their observation. Since the $uv$-plane tracks only shift marginally for consecutive nights, this sectioning by hour angle effectively corresponds to sectioning into more accurate slices of $u$ and $v$ and thus projected baselines. An arbitrary minimum observation time per hour angle can be set, leading to $m = \lfloor \frac{\mathrm{total\;observation\;time}}{\mathrm{minimum\;observation\;time }} \rfloor$ sections over which we equally distributed the total observation time. We call the resulting $4 \times m$ histograms the hour angle-sectioned cross-correlations, which were also rebinned to 10\,ps.

After rebinning, we used a Gaussian fit to center the full data and compensate for cable delay. We then averaged the channel pairs that correspond to auto- and cross-correlations to get $2 + m$ processed histograms. Each of these histograms corresponds to data taken at a baseline $r_n$. For each processed histogram, we numerically integrated $g^{(2)}(r_n, \tau) - 1$ to get a single value for $\tau_c \, V^2$. The integration used symmetric bounds covering 97.5\% of the area under the curve of a timing-resolution measurement recorded with the respective detectors and a femtosecond pulsed laser. Due to atmospheric turbulence and imperfect guiding, the RMS width of both the auto- and cross-correlations was larger than the RMS of the timing resolution measurement. If we had used the integration bounds obtained from the femtosecond pulsed laser measurements, we would have systematically underestimated our values for  $\tau_c \, V^2$ in the numerical integration. Instead we rescaled the integration boundaries by the ratio of the Gaussian fit widths of the timing resolution measurement and the respective correlation measurement. 

For the hour angle-sectioned cross-correlation, we used the integration bounds obtained from the full cross-correlation. We then rescaled the cross-correlation by $1 / \eta_\mathrm{spec} \approx 1.004$ to account for the imperfections in spectral overlap of the two interference filters. Error bars for $\tau_c\,  V^2$ follow from Gaussian error propagation of the histogram bin errors, which we take as the standard deviation of a long section of the histogram far from the peak. To validate this approach we calculated the fraction by which our calculated noise exceeds the expected shot noise, which is $0.4$\,\% for the April and  $0.2$\,\%  for the August 2024 campaign. Our data is thus very close to being shot noise limited. Error bars for the projected baselines $r_n$ were obtained as the standard deviations of the projected baselines of all $1\,s$ chunks contained within one hour angle slice. The resulting curve was fitted using Eq.~(\ref{eq:fit_func}) and orthogonal distance regression (ODR) implemented in SciPy \citep{Virtanen2020} to obtain UD diameters. The ODR fitting accounts for the error bars both in the projected baseline and in the measured coherence.

\subsection{Stellar atmosphere model}
\label{sec:saa}
\begin{figure}
    \includegraphics[width=\linewidth]{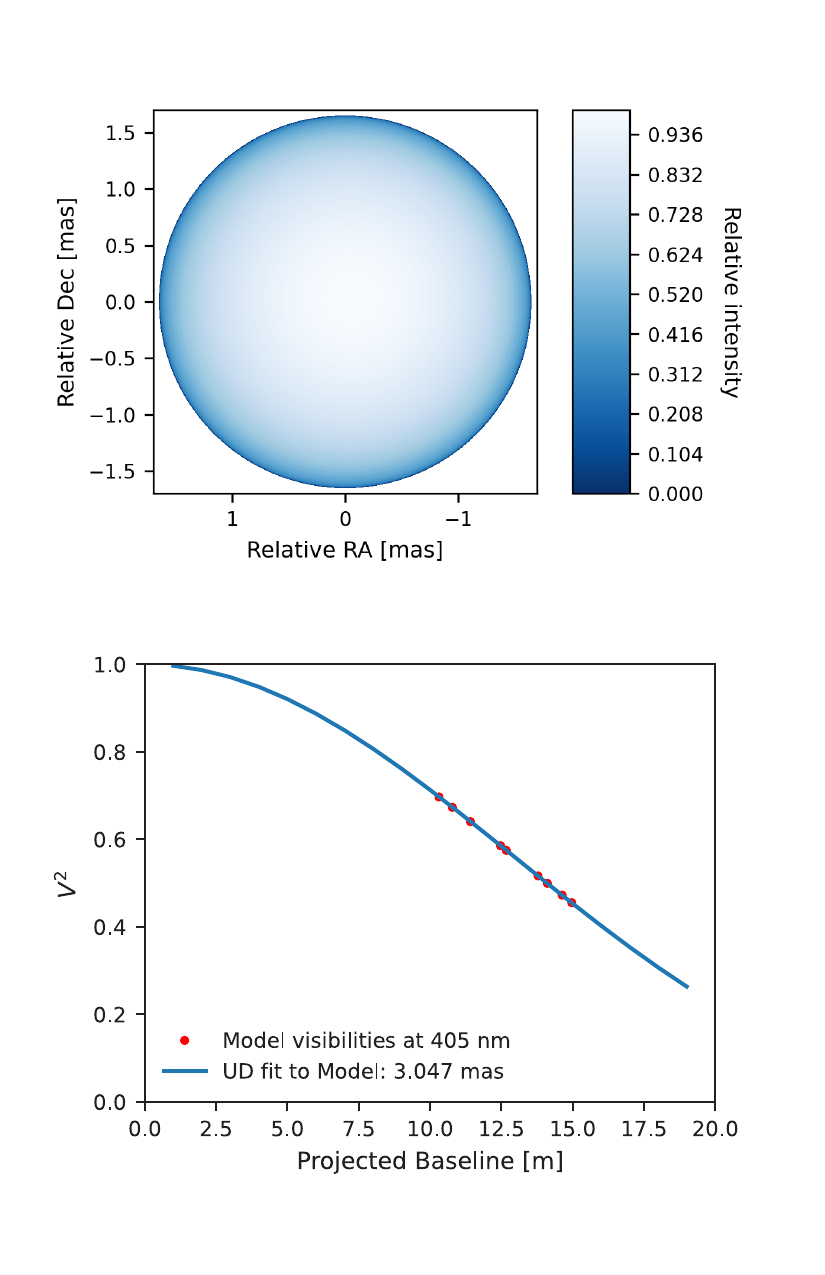}
    \caption{Synthetic intensity map for Vega at 405 nm (upper) and corresponding squared visibilities with uniform disk fit (lower).}
    \label{fig:vega_model}
\end{figure}

We computed synthetic squared visibilities for a model of Vega at 405
nm, employing stellar radiation field output from the PHOENIX code
\citep{PHOENIX_2025} and the methods of \cite{Vega_2006A}. Using the parameters from
\cite{Monnier_2012}, Vega is a nearly pole-on rapid rotator, the rotation
axis being inclined 6.2$\degr$ from the line of sight (position angle 302$\degr$), rotating at 217
km s$^{-1}$ at its equator, 77\% of its angular break-up rate.  As a
result of this rotation, Vega exhibits both limb and gravity darkening,
the latter characterized by the von-Zeipel exponent $\beta = 0.23$
which results in a temperature gradient of 1214 K between the 10700 K
pole and 9845 K equator.  A polar gravity of $\log_{10}\,g =4.1\,\text{(cgs units)}$
and mass of 2.14 M$_\odot$ correspond to a polar radius of 2.42
R$_\odot$. The true angular diameter is 3.30 mas with a parallax of 130.23
mas \citep{hipparcos}. 
With the parameters above, we computed a grid of 20 PHOENIX stellar radiation fields and interpolated these intensities to build map of Vega's photosphere on 
sky, see Figure~\ref{fig:vega_model}.  We then fit a uniform disk
model to the synthetic squared visibilities evaluated at
nine points in the $u,v$-plane. The prediction at 405 nm is that
the strongly limb and gravity darkened Vega with an angular dimeter of 
3.300 mas appears to be 3.047 mas as a uniform disk.

\section{Measurement Campaigns}\label{sec:campaign}
The first campaign took place from April 3 to April 10, 2024, using HPDs, CFDs, and the LINTag from Photonscore. We observed Vega for 10.2\,h during the nights of April 4, 5, and 6, 2024. Due to adverse weather conditions, we were unable to open the dome to observe Vega during the other nights. During the observation nights, we achieved an average count rate of 2.84\,MHz for Vega, which was in line with our expectations considering the attenuation by the atmosphere. We set the minimum observation time for hour angle-based sectioning of the projected baselines to 3\,h, resulting in $m_\mathrm{April}=3$ hour angle sections for the cross-correlation.

The second campaign took place from August 12 to August 23, 2024. The HPDs and CFDs were replaced by the PhotonPix detectors to validate their use for SII. We observed Vega for 30.3\,h. Observations were interrupted by high clouds, strong winds, and high humidity. The count rates at Epsilon were approximately 42\% lower than those observed at the Omicron telescope. This was resolved by removing the baffle of the Epsilon telescope at the end of the night of August 14, 2024. Since we also observed under non-optimal conditions, we attained only an average count rate of 2.62\,MHz, despite the nearly 54\% higher quantum efficiency of the PhotonPix detectors. We set the minimum observation time for hour angle-based sectioning of the projected baselines to 5\,h, resulting in $m_\mathrm{August} = 6$ hour angle sections for the cross-correlation.

\section{Results and Discussion}
\label{sec:results}
\subsection{Temporal auto-correlations}

\begin{figure}
    \centering
    \includegraphics[width=\linewidth]{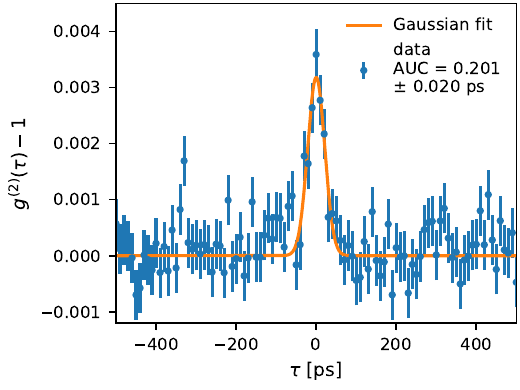}
    \includegraphics[width=\linewidth]{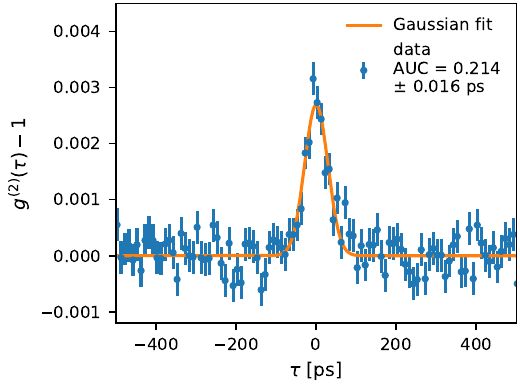}
    \caption{Temporal auto-correlation result for the April 2024 campaign (upper) and for the August 2024 campaign (lower) with corresponding error bars. The Gaussian fit used for centering and RMS estimation is shown in orange. AUC refers to the numerically integrated area under the curve of $g^{(2)}(\tau)$. The measurement result fits well to the expected coherence time of 0.21\,ps.}
    \label{fig:autoaprilaugust}
\end{figure}

Figure~\ref{fig:autoaprilaugust} shows the combined data for temporal auto-correlations measured in April and August 2024. From the narrow-band interference filter spectra, we calculated an expected coherence time of $\tau_c = 0.21$\,ps at zero-baseline with $V^2 = 1$. Both the measurement recorded in April 2024 and the measurement recorded in August 2024 reproduce this value well and lie within each other's uncertainty, see Tab.~\ref{tab:auto_corr_res_table}. 

Due to the slightly worse timing resolution of a PhotonPix compared to that of an HPD, the RMS of the correlation peak increases by a factor of $1.36$; see Tab.~\ref{tab:auto_corr_res_table}. This is disadvantageous for our visibility-recovery scheme: integrating over more temporal bins increases the noise of the area-under-the-curve integral. When comparing the S/N of the observations in April and August 2024 we insert into Eq.\,(\ref{eq:10}) to calculate the expected improvement in S/N:
\begin{equation}
    \frac{2.62 \, \mathrm{MHz}}{2.84 \, \mathrm{MHz}} \sqrt{\frac{30.3 \, \mathrm{h}}{10.2 \, \mathrm{h}} \frac{21.4 \, \mathrm{ps}}{ 29.1 \, \mathrm{ps}}} = 1.36
\end{equation}
This result agrees well with the observed 1.34 times larger S/N we measured in August when compared to the April measurement. Yet, both the timing resolution of the PhotonPix and that of the HPD lie well below the temporal resolution of current and future IACTs such as MAGIC or the CTA LSTs \citep{Karl2022}.

\begin{table}[]
    \caption{Parameters extracted from the measurement data of temporal auto-correlations shown in Fig.~\ref{fig:autoaprilaugust}.  AUC refers to the numerically integrated area under the curve of $g^{(2)}(\tau)$. The S/N shown here is the S/N of the area under the curve of  $g^{(2)}(\tau)$.}
    \begin{tabular}{c|ccc}
               & AUC  & S/N    & $\mathrm{RMS}[g^{(2)}(\tau) -1]$ \\ \hline
    Apr. 2024 &   $0.201 \pm 0.020$\,ps                &   10.0                    &     $21.4 \pm 2.2$\,ps                            \\
    Aug. 2024 &   $0.214 \pm 0.016$\,ps                &   13.4                    &     $29.1 \pm 2.0$\,ps                    
    \end{tabular}
    \label{tab:auto_corr_res_table}
\end{table}

\subsection{Cross-correlations}
\begin{figure}
    \centering
    \includegraphics[width=\linewidth]{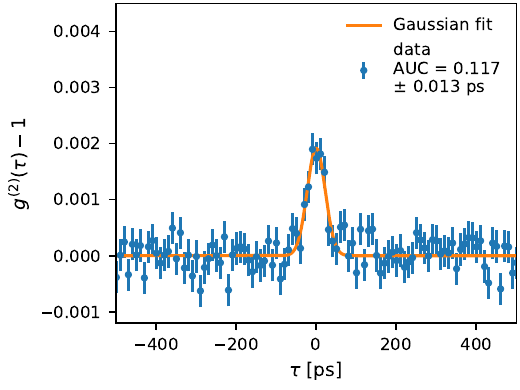}
    \hfill
    \includegraphics[width=\linewidth]{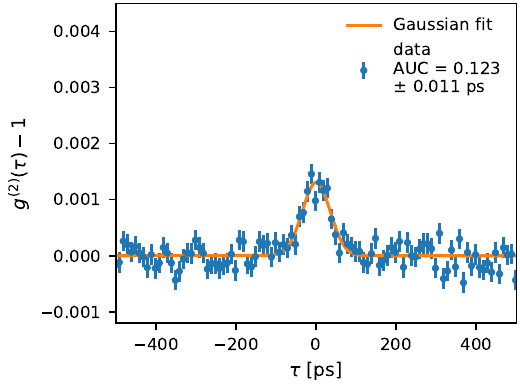}
       \caption{Cross-correlations combining all detector
pairs for the April (upper) and August (lower) 2024 campaign with corresponding error bars. The Gaussian fit used for centering and RMS estimation is shown in orange. AUC refers to the numerically integrated area under the curve of $g^{(2)}(\tau)$.}
    \label{fig:crossaprilaugust}
\end{figure}

Figure~\ref{fig:crossaprilaugust} shows the cross-correlations combining all detector pairs. From the auto-correlation results, we expect the cross-correlation measured in August 2024 to have higher S/N, but to also show worse RMS timing resolution. The values displayed in Tab.~\ref{tab:cross_corr_res_table} support these expectations. For our April 2024 measurements, however, the RMS timing resolution between the auto- and cross-correlations barely changes when comparing the results in Tab.~\ref{tab:auto_corr_res_table} and Tab.~\ref{tab:cross_corr_res_table}. For reasonably accurate tracking (approximately $10$\,arcsec between both telescopes) of Vega, this is the result we expect. In contrast, in August 2024 the auto- and cross-correlation RMSs differ by more than their $1\sigma$ error bars, though they remain compatible within $1.2\sigma$. Still, this nearly 20\% deviation deserves further investigation in future observations, as it cannot be explained by telescope tracking defects. 
 
Altogether, despite consisting of four cross-correlations rather than two auto-correlations, the cross-correlations have lower S/N than the auto-correlations. Assuming that all detectors detect equal photon rates and that the histograms are shot-noise-limited, the expected decrease in S/N can be calculated as $2/\sqrt{2} \times 1/V^2(r)$. Using the UD diameters fitted to the individual measurements in Sect.~\ref{subsec: compare_to_atms_model}, we expect the S/N of the cross-correlation measurement to be decreased by a factor of $0.82 \pm 0.09$ (April 2024) and $0.86 \pm 0.11$ (August 2024) with respect to the auto-correlation measurement, which is in agreement with our results (see Tab.~\ref{tab:auto_corr_res_table} and Tab.~\ref{tab:cross_corr_res_table}). 

Compared to previous photon-counting SII measurements with RMS timing resolutions of $\ge 212$\,ps \citep{Guerin2017, Zampieri2021, Horch2022, Mozdzen2025}, our results improve the timing resolution by a factor of $> 5$. Furthermore, our measurements are the first spatial SII measurements using single-photon-counting detectors at the blue wavelength typically used by SII instruments at IACTs.

\begin{table}[]
    \caption{Parameters extracted from the measurement data of cross-correlations shown in Fig.~\ref{fig:crossaprilaugust}.  AUC refers to the numerically integrated area under the curve of $g^{(2)}(\tau)$. The S/N shown here is the S/N of the area under the curve of  $g^{(2)}(\tau)$.}
    \begin{tabular}{c|ccc}
               & AUC  & S/N    & $\mathrm{RMS}[g^{(2)}(\tau) -1]$ \\ \hline
    Apr. 2024 &   $0.117 \pm 0.013$\,ps                &   9.2                    &     $22.7 \pm 2.4$\,ps                            \\
    Aug. 2024 &   $0.123 \pm 0.011$\,ps                &   11.2                   &     $34.8 \pm 2.8$\,ps                           
    \end{tabular}
       
    \label{tab:cross_corr_res_table}
\end{table}

\subsection{Comparison to a stellar atmosphere model for Vega}
\label{subsec: compare_to_atms_model}

\begin{figure}
    \centering
    \includegraphics[trim={0cm 1cm 0cm 1cm},clip, width=0.8\linewidth]{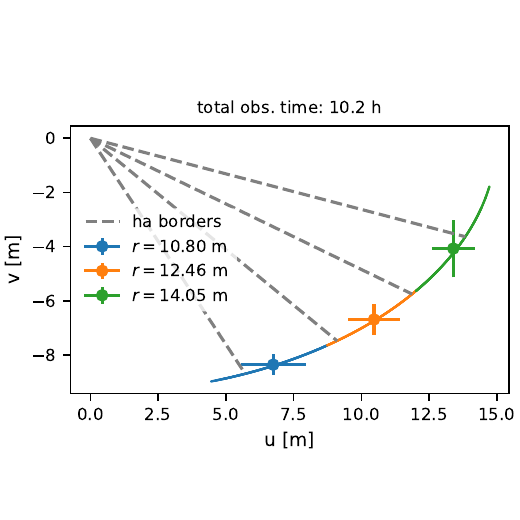}
    \includegraphics[trim={0cm 0.5cm 0cm 0.5cm},clip, width=0.8\linewidth]{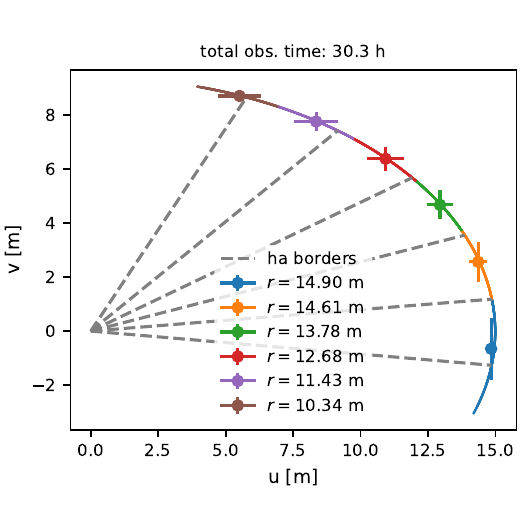}
        \caption{Average $uv$ coordinates and total $uv$-plane coverage of the April (upper) and August (lower) 2024 campaign. Different slices of the hour angle are marked in different colors, and the average $uv$ coordinates of each slice are shown with their respective error bars and average projected baselines. Since the measurement is sliced so that each slice has equal observation time, the slices contain different hour angles and do not end at the hour angle boundaries shown as dashed gray lines. For faster rendering, and since each 1\,s chunk produces a single $uv$ point, we plot only every 25th $uv$ point to form the continuous line.}

    \label{fig:uvplaneaprilaugust}
\end{figure}

Fitting directly to the total auto- and cross-correlations is possible, but the fit error would then be limited by the large projected-baseline error. We thus separated our observations into equal observation time hour angle sections, namely $m = 3$ hour angle sections for the April measurement and $m = 6$ hour angle sections for the August measurement (see Sec.~\ref{sec:campaign}), which gives a smaller projected baseline error—and thus a more accurate fit. Fig.~\ref{fig:uvplaneaprilaugust} shows the $uv$-plane coverage of the recorded histograms, the hour angle sections, and the average $u$ and $v$ coordinates within these sections. Note that due to the difference in observation month, Vega follows a different track in the $uv$-plane for our two measurement campaigns.

\begin{figure*}
    \centering
    \includegraphics[width=0.33\linewidth]{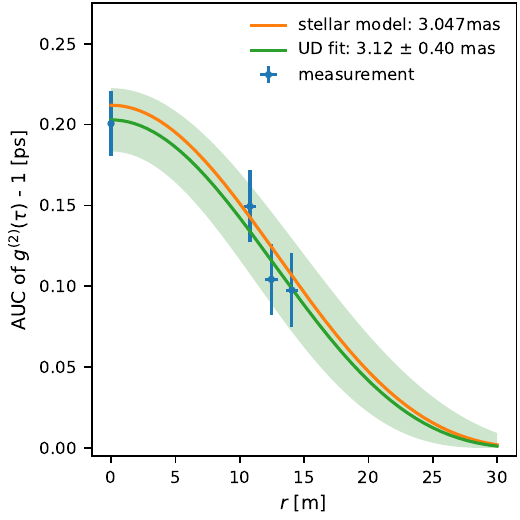}
    \includegraphics[width=0.33\linewidth]{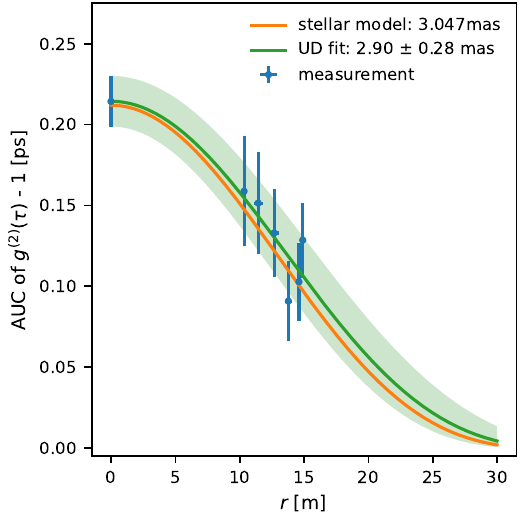}
    \includegraphics[width=0.33\linewidth]{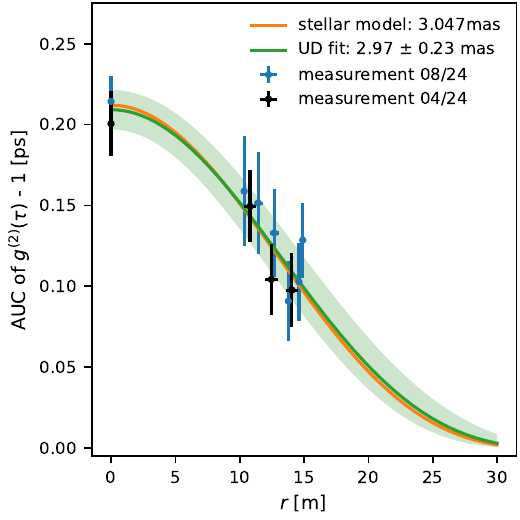}
        \caption{Visibility curve for the April (left), August (middle) 2024 campaign, and the combined data (right). The April campaign used HPDs and CFDs from Photonscore, whereas the August 2024 campaign used the PhotonPix detector. The cross-correlation data were rebinned according to the hour angle as shown in Fig.~\ref{fig:uvplaneaprilaugust}. In the right panel, data points in black correspond to the campaign in April 2024 and blue data points correspond to the campaign in August 2024. The UD fit for the combined data (right) yields an angular diameter of $2.97\pm0.23$\,mas for Vega, which is in agreement with the stellar atmosphere model.}

    \label{fig:v2combined}
\end{figure*}

In Fig.~\ref{fig:v2combined}, we fit the UD model to the hour angle sectioned visibility curves measured in April and August 2024, respectively. Using the same $uv$-plane points, we fit a UD model to the results of our stellar atmosphere model (see also Sec.~\ref{sec:saa}) to generate our baseline expectation. For Vega at 405\,nm, this results in an angular diameter of 3.047\,mas. Vega is a rapid rotator observed nearly pole-on, so we expect identical visibilities in all quadrants of the $uv$-plane; fitting a single UD diameter should therefore converge well. Both the measurement taken with the HPDs in April 2024 and the measurement taken with PhotonPix detectors in August 2024 reproduce our expectation within their error bars, at 3.12 $\pm$ 0.40\,mas and 2.90 $\pm$ 0.28\,mas fitted UD diameter, respectively. Combining the April and August 2024 data, we obtain a UD diameter of 2.97 $\pm$ 0.23\,mas (Fig.~\ref{fig:v2combined}), which agrees with our stellar-atmosphere-model expectation at an S/N of 12.9. This result was obtained in 40.5\,h of total observation time.

\section{High time resolution SII with single-photon detectors at IACT arrays?}
\label{sec:sii_at_iacts}
Recently \cite{Cortina2025} have introduced a possible upgrade to existing SII instruments at IACTs with active mirror control (AMC) and parabolic mirrors like the CTA LSTs. Their Butterfly's eye approach segments the large, high-focal-ratio IACT mirrors into smaller sub-mirrors, each made of a few existing facets and illuminating its own single-photon detector. This is beneficial because it reduces both the timing anisochronicity and the focal ratio when compared to the full telescope dish. For segmentation into more than 7 sub-mirrors, anisochronicities of $60\,$ps FWHM ($\approx 25.5\,$ps RMS) can be reached, and $5$\,nm FWHM instead of the more conventional $> 20\,$nm FWHM optical filters can be used. \cite{Cortina2025} note that this parameter change makes the use of single-photon detectors with large area and high time resolution, like the Becker\&Hickl HPDs or the PhotonPix, feasible and more importantly beneficial to all SII science cases.

\begin{table}[]
    \caption{List of telescope parameters for the C2PU telescopes and the four CTA north site LSTs in a 7-detector Butterfly's eye configuration equipped with PhotonPix detectors. Parameters for the LSTs were taken from \cite{Cortina2025} and \cite{Abe2024}. $A$ is the primary mirror area,  $\epsilon_\mathrm{opt}$ is the product of all optical efficiencies, $\Delta t_\mathrm{opt}$ and $\Delta t_\mathrm{det}$ are the RMS widths of the individual telescope anisochronicities and two-detector timing resolution, and the $\sqrt{n_\mathrm{corr}}$ factor accounts for the multiple independent baselines measurable between the four LSTs but not between the two C2PU telescopes.}
    \begin{tabular}{c|ccccc}
          & $A$ [$\mathrm{m^2}$] & $\epsilon_\mathrm{opt}$ &  $\Delta t_\mathrm{opt}$ [ps] & $\Delta t_\mathrm{det} [ps]$ &  $\sqrt{n_\mathrm{corr}}$ \\ \hline
    C2PU  & 0.85 & 0.305 & - & 34.8 & 1 \\
    LSTs  & 370 & 0.6 & 25.5 & 34.8 & $\sqrt{6}$
    \label{tab:parameter_table}
\end{tabular}
\end{table}

\begin{figure}
    \centering
    \includegraphics{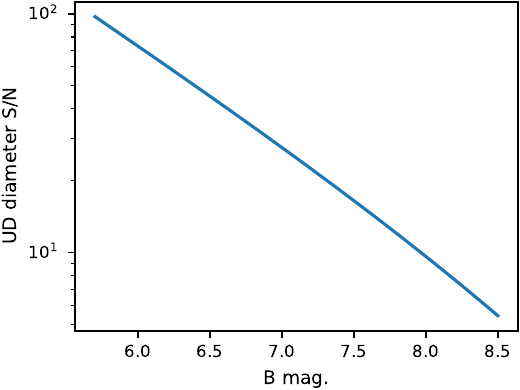}
    \caption{Expected S/N of the UD diameter measurement of a star with a given B-band magnitude observed for 20\,h with 4 LSTs equipped with a 7-detector Butterfly's eye cameras based on the PhotonPix detector.}
    \label{fig:snr_over_mag}
\end{figure}

In the following we use our stellar diameter results of Vega together with telescope parameters of the CTA LSTs taken from \cite{Cortina2025} and \cite{Abe2024} to independently gauge the possible relative uncertainty on the UD of a star at different apparent magnitudes using a 7-detector Butterfly's eye configuration equipped with PhotonPix detectors. Tab. \ref{tab:parameter_table} summarizes the parameters we use for our calculation. Further we assume that the night sky background can be reduced to approximately $9.5$\,mag. at the LSTs under dark sky conditions as discussed in \cite{Cortina2005} and that the star has similar spatial visibilities at the $100$\,m  baselines  of the CTA LSTs as Vega has at the $15$\,m baseline of C2PU. The latter criterion corresponds to a star with angular diameter $0.45$\,mas. For our estimation we start with the UD S/N and measurement time obtained during the April 2024 observation, but still assume that the LSTs are equipped with PhotonPix detectors. We assume PhotonPix detectors for two reasons. First, their larger active area better fits the constraints at the IACTs. Second, their higher quantum efficiency should outweigh their slightly worse timing resolution and let them outperform the HPDs — which they would have in August 2024 had the observing conditions been better. To avoid count-rate loss from detector dead time, we limit ourselves to detected rates below $30$\,MHz \cite{Karl2025_pub}, which for a 7-detector Butterfly's eye configuration with $5$\,nm filter bandwidth corresponds to apparent magnitudes above $5.7$. Following the treatment in \cite{Cortina2025} we calculate the expected S/N of the target's UD diameter after an observation time of $20$\,h according to Eq.\,~(\ref{eq:10}). The results are plotted in Fig.~\ref{fig:snr_over_mag}. Our estimations show that under the conditions of this simulation we expect to be able to measure the angular diameter of a mag. 7.8 star to an S/N of 10.0 in $20$\,h. For a mag. 7.8 star, the detected rate per detector with the Butterfly's eye camera at the LSTs would be about 3 MHz, compatible with our measured rates of Vega, though at 2.5 times the optical bandwidth. Our expected S/N is lower than the expected S/N of approx. $22$ reported by \cite{Cortina2025}. Since our results are based on past observation runs and taking into account that the PhotonPix has better RMS timing resolution but lower quantum efficiency than the detectors considered in their study, this result is expected.

The practical implementation of such a Butterfly's eye camera, however, would place rigid requirements on the accuracy of the 3D-positioning and guiding of the LSTs. Allowing a 10\% sensitivity drop from guiding and positioning jitter, split equally between the two telescopes, requires knowing each telescope's 3D position to $0.55$\,cm RMS and tracking the star with an RMS guiding accuracy of $0.19$ arcmin. The necessary positioning accuracy can be achieved using modern total stations, which enable sub-mm accurate position measurements that can be performed live if needed \cite{Loesler2013}, but the required pointing accuracy is below the LSTs $1$\,arcmin tracking accuracy \citep{Dalchenko2023}. The necessary pointing accuracy could be achieved after post-processing \citep{Foffano2021}, which would necessitate storing and reprocessing at least intermediate histograms.

\section{Conclusion}

We have demonstrated stellar intensity interferometry measurements of Vega at 405\,nm with sub-35\,ps RMS timing resolution, achieving the highest temporal precision reported for spatial SII to date. Using two different single-photon-counting detector types, hybrid photodetectors with 22.7\,ps timing resolution and the newly developed PhotonPix detectors with 34.8\,ps timing resolution, we measured angular diameters of $3.12 \pm 0.40$\,mas and $2.90 \pm 0.28$\,mas in April and August 2024, respectively. Combining both campaigns yields a diameter of $2.97 \pm 0.23$\,mas with an S/N ratio of 12.9, obtained in 40.5\,h total observation time. These measurements agree with stellar atmosphere model predictions taking into account  both limb and gravity darkening, validating both our methodology and the PhotonPix detector technology for precision astronomical applications.

Our results represent a factor $>5$ improvement in timing resolution compared to previous photon-counting SII campaigns and constitute the first spatial SII measurements using single-photon-counting detectors at blue wavelengths (405\,nm) employed by IACT-based SII instruments. We have validated the use of White Rabbit technology for UTC synchronization, enabling accurate optical path difference corrections and supporting future extension to longer baselines and multi-site configurations. We have also successfully tested a new large-area detector, the PhotonPix, and validated its use in an SII instrument. Looking toward implementation at IACTs, we confirmed that a Butterfly's eye camera equipped with only seven PhotonPix or HPD detectors is feasible from the photon detection point of view. Such an instrument, if deployed atht the 4 CTA LSTs, would allow the measurement of the UD diameter of stars up to mag. 7.8 with sn SNR of $\ge$ 10 in observation times of around 20\,h as extrapolated from our results. We however note that highly accurate telescope positioning and star tracking are necessary prerequisites for the successful implementation of a Butterfly's eye camera instrument, as is adequate control of the night sky background even during dark time.

\section*{Data Availability}
The data directly supporting the plots are available at \url{https://doi.org/10.22000/u7wqh6nus0pd9wpz}. The raw photon event stream can be supplied upon reasonable request from the corresponding authors.

\begin{acknowledgements}
We cordially thank the team of the Observatoire de la Côte d'Azur for letting us perform measurements at their telescopes. We acknowledge the Institut Géographique National (IGN) for the precise geodetic measurements of the telescope positions. This work made use of the High Performance Cluster "Vega" at Embry-Riddle Aeronautical University.
\end{acknowledgements}

%

\bibliographystyle{aa}
\bibliography{example}

@Article{HanburyBrown1956,
  author  = {{Hanbury Brown}, R and Twiss, R Q},
  journal = {Nature},
  title   = {A Test of a New Type of Stellar Interferometer on Sirius},
  year    = {1956},
  month   = nov,
  number  = {4541},
  pages   = {1046--1048},
  volume  = {178},
}

@Article{Michelson1921,
  author   = {Michelson, A A and Pease, F G},
  journal  = {Proc Natl Acad Sci U S A},
  title    = {Measurement of the Diameter of {Alpha-Orionis} by the Interferometer},
  year     = {1921},
  month    = may,
  number   = {5},
  pages    = {143--146},
  volume   = {7},
  address  = {United States},
  language = {en},
}

@Article{Brown1967,
  author   = {{Hanbury Brown}, R. and Davis, J. and Allen, L. R.},
  journal  = {Monthly Notices of the Royal Astronomical Society},
  title    = {The Stellar Interferometer at Narrabri Observatory—I: A Description of The Instrument and the Observational Procedure},
  year     = {1967},
  issn     = {0035-8711},
  month    = {12},
  number   = {4},
  pages    = {375-392},
  volume   = {137},
  doi      = {10.1093/mnras/137.4.375},
  eprint   = {https://academic.oup.com/mnras/article-pdf/137/4/375/8075225/mnras137-0375.pdf},
  url      = {https://doi.org/10.1093/mnras/137.4.375},
}

@Article{HanburyBrown1974a,
  author   = {{Hanbury Brown}, R. and Davis, J. and Allen, L. R.},
  journal  = {Monthly Notices of the Royal Astronomical Society},
  title    = {{The Angular Diameters of 32 Stars}},
  year     = {1974},
  issn     = {0035-8711},
  month    = {04},
  number   = {1},
  pages    = {121-136},
  volume   = {167},
  doi      = {10.1093/mnras/167.1.121},
  eprint   = {https://academic.oup.com/mnras/article-pdf/167/1/121/8079851/mnras167-0121.pdf},
  url      = {https://doi.org/10.1093/mnras/167.1.121},
}

@Article{Labeyrie1975,
  author   = {{Labeyrie}, A.},
  journal  = {The Astrophysical Journal Letters},
  title    = {{Interference fringes obtained on Vega with two optical telescopes.}},
  year     = {1975},
  month    = mar,
  pages    = {L71-L75},
  volume   = {196},
  adsurl   = {https://ui.adsabs.harvard.edu/abs/1975ApJ...196L..71L},
  doi      = {10.1086/181747},
}

@Article{GRAVITY+Collaboration2022,
  author  = {{GRAVITY+ Collaboration} and {Abuter, R.} and {Allouche, F.} and {Amorim, A.} and {Bailet, C.} and {Bauböck, M.} and {Berger, J.-P.} and {Berio, P.} and {Bigioli, A.} and {Boebion, O.} and {Bolzer, M. L.} and {Bonnet, H.} and {Bourdarot, G.} and {Bourget, P.} and {Brandner, W.} and {Clénet, Y.} and {Courtney-Barrer, B.} and {Dallilar, Y.} and {Davies, R.} and {Defrère, D.} and {Delboulbé, A.} and {Delplancke, F.} and {Dembet, R.} and {de Zeeuw, P. T.} and {Drescher, A.} and {Eckart, A.} and {Édouard, C.} and {Eisenhauer, F.} and {Fabricius, M.} and {Feuchtgruber, H.} and {Finger, G.} and {Förster Schreiber, N. M.} and {Garcia, E.} and {Garcia, P.} and {Gao, F.} and {Gendron, E.} and {Genzel, R.} and {Gil, J. P.} and {Gillessen, S.} and {Gomes, T.} and {Gonté, F.} and {Gouvret, C.} and {Guajardo, P.} and {Guieu, S.} and {Hartl, M.} and {Haubois, X.} and {Haußmann, F.} and {Heißel, G.} and {Henning, Th.} and {Hippler, S.} and {Hönig, S.} and {Horrobin, M.} and {Hubin, N.} and {Jacqmart, E.} and {Jochum, L.} and {Jocou, L.} and {Kaufer, A.} and {Kervella, P.} and {Korhonen, H.} and {Kreidberg, L.} and {Lacour, S.} and {Lagarde, S.} and {Lai, O.} and {Lapeyrère, V.} and {Laugier, R.} and {Le Bouquin, J.-B.} and {Leftley, J.} and {Léna, P.} and {Lutz, D.} and {Mang, F.} and {Marcotto, A.} and {Maurel, D.} and {Mérand, A.} and {Millour, F.} and {More, N.} and {Nowacki, H.} and {Nowak, M.} and {Oberti, S.} and {Ott, T.} and {Pallanca, L.} and {Pasquini, L.} and {Paumard, T.} and {Perraut, K.} and {Perrin, G.} and {Petrov, R.} and {Pfuhl, O.} and {Pourré, N.} and {Rabien, S.} and {Rau, C.} and {Robbe-Dubois, S.} and {Rochat, S.} and {Salman, M.} and {Schöller, M.} and {Schubert, J.} and {Schuhler, N.} and {Shangguan, J.} and {Shimizu, T.} and {Scheithauer, S.} and {Sevin, A.} and {Soulez, F.} and {Spang, A.} and {Stadler, E.} and {Stadler, J.} and {Straubmeier, C.} and {Sturm, E.} and {Tacconi, L. J.} and {Tristram, K. R. W.} and {Vincent, F.} and {von Fellenberg, S.} and {Uysal, S.} and {Widmann, F.} and {Wieprecht, E.} and {Wiezorrek, E.} and {Woillez, J.} and {Yazici, S.} and {Young, A.} and {Zins, G.}},
  journal = {A\&A},
  title   = {First light for GRAVITY Wide - Large separation fringe tracking for the Very Large Telescope Interferometer},
  year    = {2022},
  pages   = {A75},
  volume  = {665},
  doi     = {10.1051/0004-6361/202243941},
  url     = {https://doi.org/10.1051/0004-6361/202243941},
}

@Article{Anugu2020,
  author    = {Anugu, Narsireddy and Le Bouquin, Jean-Baptiste and Monnier, John D. and Kraus, Stefan and Setterholm, Benjamin R. and Labdon, Aaron and Davies, Claire L and Lanthermann, Cyprien and Gardner, Tyler and Ennis, Jacob and Johnson, Keith J. C. and Ten Brummelaar, Theo and Schaefer, Gail and Sturmann, Judit},
  journal   = {The Astronomical Journal},
  title     = {MIRC-X: A Highly Sensitive Six-telescope Interferometric Imager at the CHARA Array},
  year      = {2020},
  month     = {sep},
  number    = {4},
  pages     = {158},
  volume    = {160},
  doi       = {10.3847/1538-3881/aba957},
  publisher = {The American Astronomical Society},
  url       = {https://doi.org/10.3847/1538-3881/aba957},
}

@Article{Dravins2013,
  author   = {Dainis Dravins and Stephan LeBohec and Hannes Jensen and Paul D. Nuñez},
  journal  = {Astroparticle Physics},
  title    = {Optical intensity interferometry with the Cherenkov Telescope Array},
  year     = {2013},
  issn     = {0927-6505},
  note     = {Seeing the High-Energy Universe with the Cherenkov Telescope Array - The Science Explored with the CTA},
  pages    = {331-347},
  volume   = {43},
  doi      = {https://doi.org/10.1016/j.astropartphys.2012.04.017},
  url      = {https://www.sciencedirect.com/science/article/pii/S0927650512001314},
}

@Article{Guerin2018,
  author   = {Guerin, W and Rivet, J-P and Fouché, M and Labeyrie, G and Vernet, D and Vakili, F and Kaiser, R},
  journal  = {Monthly Notices of the Royal Astronomical Society},
  title    = {{Spatial intensity interferometry on three bright stars}},
  year     = {2018},
  issn     = {0035-8711},
  month    = {07},
  number   = {1},
  pages    = {245-250},
  volume   = {480},
  doi      = {10.1093/mnras/sty1792},
  eprint   = {https://academic.oup.com/mnras/article-pdf/480/1/245/25230273/sty1792.pdf},
  url      = {https://doi.org/10.1093/mnras/sty1792},
}

@Article{Abeysekara2020,
  author   = {Abeysekara, A. U. and Benbow, W. and Brill, A. and Buckley, J. H. and Christiansen, J. L. and Chromey, A. J. and Daniel, M. K. and Davis, J. and Falcone, A. and Feng, Q. and Finley, J. P. and Fortson, L. and Furniss, A. and Gent, A. and Giuri, C. and Gueta, O. and Hanna, D. and Hassan, T. and Hervet, O. and Holder, J. and Hughes, G. and Humensky, T. B. and Kaaret, P. and Kertzman, M. and Kieda, D. and Krennrich, F. and Kumar, S. and LeBohec, T. and Lin, T. T. Y. and Lundy, M. and Maier, G. and Matthews, N. and Moriarty, P. and Mukherjee, R. and Nievas-Rosillo, M. and O'Brien, S. and Ong, R. A. and Otte, A. N. and Pfrang, K. and Pohl, M. and Prado, R. R. and Pueschel, E. and Quinn, J. and Ragan, K. and Reynolds, P. T. and Ribeiro, D. and Richards, G. T. and Roache, E. and Ryan, J. L. and Santander, M. and Sembroski, G. H. and Wakely, S. P. and Weinstein, A. and Wilcox, P. and Williams, D. A. and Williamson, T. J.},
  journal  = {Nature Astronomy},
  title    = {Demonstration of stellar intensity interferometry with the four VERITAS telescopes},
  year     = {2020},
  issn     = {2397-3366},
  month    = {Dec},
  number   = {12},
  pages    = {1164-1169},
  volume   = {4},
  day      = {01},
  doi      = {10.1038/s41550-020-1143-y},
  url      = {https://doi.org/10.1038/s41550-020-1143-y},
}

@Article{Acciari2019a,
  author   = {Acciari, V A and Bernardos, M I and Colombo, E and Contreras, J L and Cortina, J and De Angelis, A and Delgado, C and Díaz, C and Fink, D and Mariotti, M and Mangano, S and Mirzoyan, R and Polo, M and Schweizer, T and Will, M},
  journal  = {Monthly Notices of the Royal Astronomical Society},
  title    = {Optical intensity interferometry observations using the MAGIC Imaging Atmospheric Cherenkov Telescopes},
  year     = {2019},
  issn     = {0035-8711},
  month    = {11},
  number   = {2},
  pages    = {1540-1547},
  volume   = {491},
  doi      = {10.1093/mnras/stz3171},
  eprint   = {https://academic.oup.com/mnras/article-pdf/491/2/1540/31144895/stz3171.pdf},
  url      = {https://doi.org/10.1093/mnras/stz3171},
}

@Article{Zmija2023,
  author   = {Zmija, Andreas and Vogel, Naomi and Wohlleben, Frederik and Anton, Gisela and Zink, Adrian and Funk, Stefan},
  journal  = {Monthly Notices of the Royal Astronomical Society},
  title    = {First intensity interferometry measurements with the H.E.S.S. telescopes},
  year     = {2023},
  issn     = {0035-8711},
  month    = {12},
  number   = {4},
  pages    = {12243-12252},
  volume   = {527},
  doi      = {10.1093/mnras/stad3676},
  eprint   = {https://academic.oup.com/mnras/article-pdf/527/4/12243/55568198/stad3676.pdf},
  url      = {https://doi.org/10.1093/mnras/stad3676},
}

@Article{Abe2024,
  author   = {Abe, S and Abhir, J and Acciari, V A and Aguasca-Cabot, A and Agudo, I and Aniello, T and Ansoldi, S and Antonelli, L A and Arbet Engels, A and Arcaro, C and Artero, M and Asano, K and Babić, A and Baquero, A and de Almeida, U Barres and Barrio, J A and Batković, I and Bautista, A and Baxter, J and González, J Becerra and Bernardini, E and Bernardos, M and Bernete, J and Berti, A and Besenrieder, J and Bigongiari, C and Biland, A and Blanch, O and Bonnoli, G and Bošnjak, Ž and Burelli, I and Busetto, G and Campoy-Ordaz, A and Carosi, A and Carosi, R and Carretero-Castrillo, M and Ceribella, G and Chai, Y and Cifuentes, A and Colombo, E and Contreras, J L and Cortina, J and Covino, S and D’Amico, G and D’Elia, V and Da Vela, P and Dazzi, F and De Angelis, A and De Lotto, B and de Menezes, R and Del Popolo, A and Delfino, M and Delgado, J and Mendez, C Delgado and Pierro, F Di and Venere, L Di and Prester, D Dominis and Donini, A and Dorner, D and Doro, M and Elsaesser, D and Emery, G and Escudero, J and Fariña, L and Fattorini, A and Foffano, L and Font, L and Fröse, S and Fukami, S and Fukazawa, Y and López, R J García and Garczarczyk, M and Gasparyan, S and Gaug, M and Paiva, J G Giesbrecht and Giglietto, N and Giordano, F and Gliwny, P and Gradetzke, T and Grau, R and Green, D and Green, J G and Günther, P and Hadasch, D and Hahn, A and Hassan, T and Heckmann, L and Herrera, J and Hrupec, D and Hütten, M and Imazawa, R and Ishio, K and Jiménez Martínez, I and Jormanainen, J and Kayanoki, T and Kerszberg, D and Kluge, G W and Kobayashi, Y and Kouch, P M and Kubo, H and Kushida, J and Láinez, M and Lamastra, A and Leone, F and Lindfors, E and Linhoff, L and Lombardi, S and Longo, F and López-Coto, R and López-Moya, M and López-Oramas, A and Loporchio, S and Lorini, A and Lyard, E and Fraga, B Machado de Oliveira and Majumdar, P and Makariev, M and Maneva, G and Mang, N and Manganaro, M and Mangano, S and Mannheim, K and Mariotti, M and Martínez, M and Martínez-Chicharro, M and Mas-Aguilar, A and Mazin, D and Menchiari, S and Mender, S and Miceli, D and Miener, T and Miranda, J M and Mirzoyan, R and González, M Molero and Molina, E and Mondal, H A and Moralejo, A and Morcuende, D and Nakamori, T and Nanci, C and Neustroev, V and Nickel, L and Rosillo, M Nievas and Nigro, C and Nikolić, L and Nilsson, K and Nishijima, K and Ekoume, T Njoh and Noda, K and Nozaki, S and Ohtani, Y and Okumura, A and Otero-Santos, J and Paiano, S and Palatiello, M and Paneque, D and Paoletti, R and Paredes, J M and Peresano, M and Persic, M and Pihet, M and Pirola, G and Podobnik, F and Moroni, P G Prada and Prandini, E and Principe, G and Priyadarshi, C and Rhode, W and Ribó, M and Rico, J and Righi, C and Sahakyan, N and Saito, T and Satalecka, K and Saturni, F G and Schleicher, B and Schmidt, K and Schmuckermaier, F and Schubert, J L and Schweizer, T and Sciaccaluga, A and Silvestri, G and Sitarek, J and Sliusar, V and Sobczynska, D and Spolon, A and Stamerra, A and Strišković, J and Strom, D and Strzys, M and Suda, Y and Surić, T and Suutarinen, S and Tajima, H and Takahashi, M and Takeishi, R and Temnikov, P and Terauchi, K and Terzić, T and Teshima, M and Truzzi, S and Tutone, A and Ubach, S and van Scherpenberg, J and Acosta, M Vazquez and Ventura, S and Viale, I and Vigorito, C F and Vitale, V and Walter, R and Will, M and Wunderlich, C and Yamamoto, T and Chon, G and Díaz, C and Fiori, M and Lobo, M and Naletto, G and Polo, M and Rodríguez-Vázquez, J J and Saha, P and Zampieri, L},
  journal  = {Monthly Notices of the Royal Astronomical Society},
  title    = {Performance and first measurements of the MAGIC stellar intensity interferometer},
  year     = {2024},
  issn     = {0035-8711},
  month    = {03},
  number   = {4},
  pages    = {4387-4404},
  volume   = {529},
  doi      = {10.1093/mnras/stae697},
  eprint   = {https://academic.oup.com/mnras/article-pdf/529/4/4387/57147987/stae697.pdf},
  url      = {https://doi.org/10.1093/mnras/stae697},
}

@Misc{Kieda2021a,
  author        = {David Kieda and Jonathan Davis and Tugdual LeBohec and Mike Lisa and Nolan K. Matthews},
  title         = {The VERITAS-Stellar Intensity Interferometry (VSII) survey of Stellar Diameters},
  year          = {2021},
  archiveprefix = {arXiv},
  eprint        = {2108.09774},
  primaryclass  = {astro-ph.SR},
}

@Article{Vogel2024,
  author   = {Vogel, Naomi and Zmija, Andreas and Wohlleben, Frederik and Anton, Gisela and Mitchell, Alison and Zink, Adrian and Funk, Stefan},
  journal  = {Monthly Notices of the Royal Astronomical Society},
  title    = {Simultaneous two-colour intensity interferometry with H.E.S.S},
  year     = {2024},
  issn     = {0035-8711},
  month    = {11},
  number   = {3},
  pages    = {2334-2341},
  volume   = {537},
  doi      = {10.1093/mnras/stae2643},
  eprint   = {https://academic.oup.com/mnras/article-pdf/537/3/2334/60828614/stae2643.pdf},
  url      = {https://doi.org/10.1093/mnras/stae2643},
}

@InProceedings{Karl2025_pub,
  author       = {Sebastian Karl and Verena Leopold and Stefan Richter and Yury Prokazv and Evgeny Turbin and Gennady Sintotskiy and Dimitry Orlov and Joachim {von Zanthier}},
  booktitle    = {SPIE Defense and Commercial Sensing 2025},
  title        = {{High-detection efficiency MCP-PMTs with single photon counting capability for LIDAR applications}},
  year         = {2025},
  editor       = {Zoran Sodnik and Nikos Karafolas and Bruno Cugny},
  organization = {International Society for Optics and Photonics},
  pages        = {666},
  publisher    = {SPIE},
  volume       = {666},
  doi          = {10.1117/12.2536028},
  url          = {https://doi.org/10.1117/12.2536028},
}

@Article{Leopold2025,
  author    = {Verena G. Leopold and Sebastian Karl and Jean-Pierre Rivet and Joachim von Zanthier},
  journal   = {Journal of Astronomical Telescopes, Instruments, and Systems},
  title     = {{On-sky demonstration of an ultra-fast intensity interferometry instrument utilizing hybrid single photon counting detectors}},
  year      = {2025},
  number    = {3},
  pages     = {035005},
  volume    = {11},
  doi       = {10.1117/1.JATIS.11.3.035005},
  publisher = {SPIE},
  url       = {https://doi.org/10.1117/1.JATIS.11.3.035005},
}

@Book{HanburyBrown1974,
  author    = {Hanbury Brown, Robert},
  publisher = {Taylor \& Francis, Halsted Press},
  title     = {The intensity interferometer; its application to astronomy},
  year      = {1974},
  isbn      = {9780470107973,0470107979,9780850660722,0850660726},
  url       = {http://gen.lib.rus.ec/book/index.php?md5=d604cdb5cacf1ea40203b4b71256ea19},
}

@Book{Fox2006,
  author    = {Mark Fox},
  publisher = {Oxford University Press, USA},
  title     = {Quantum optics: an introduction},
  year      = {2006},
  isbn      = {0198566735,9780198566731,9781435606807},
  series    = {Oxford Master Series in Physics, 6},
  url       = {http://gen.lib.rus.ec/book/index.php?md5=AF074928C421227E6B7EBFCC30425DE2},
}

@Book{Siegert1943,
  author    = {Siegert, AJF},
  publisher = {Radiation Laboratory, Massachusetts Institute of Technology},
  title     = {On the fluctuations in signals returned by many independently moving scatterers},
  year      = {1943},
}

@Book{Mandel1995,
  author    = {Mandel, Leonard and Wolf, Emil},
  publisher = {Cambridge University Press},
  title     = {Optical Coherence and Quantum Optics},
  year      = {1995},
  doi       = {10.1017/CBO9781139644105},
  place     = {Cambridge},
}

@Book{Labeyrie2006,
  author    = {Labeyrie, A. and Lipson, S. G. and Nisenson, P.},
  publisher = {Cambridge University Press},
  title     = {An Introduction to Optical Stellar Interferometry},
  year      = {2006},
  address   = {Cambridge},
  isbn      = {9780521828727},
  doi       = {10.1017/CBO9780511617638},
  url       = {https://www.cambridge.org/core/books/an-introduction-to-optical-stellar-interferometry/2EA3ABDA8557CF3277063391C02E899D},
}

@Misc{Rhodes2019,
  author       = {Rhodes, Brandon},
  howpublished = {\url{https://ui.adsabs.harvard.edu/abs/2019ascl.soft07024R}},
  note         = {Astrophysics Source Code Library, ascl:1907.024},
  title        = {{Skyfield}: Generate high precision research-grade positions for stars, planets, moons, and Earth satellites},
  year         = {2019},
}

@Book{Thompson2017,
  author    = {Thompson, A Richard and Moran, James M and Swenson, George W},
  publisher = {Springer Nature},
  title     = {Interferometry and synthesis in radio astronomy},
  year      = {2017},
}

@Article{Virtanen2020,
  author  = {Virtanen, Pauli and Gommers, Ralf and Oliphant, Travis E. and Haberland, Matt and Reddy, Tyler and Cournapeau, David and Burovski, Evgeni and Peterson, Pearu and Weckesser, Warren and Bright, Jonathan and {van der Walt}, St{\'e}fan J. and Brett, Matthew and Wilson, Joshua and Millman, K. Jarrod and Mayorov, Nikolay and Nelson, Andrew R. J. and Jones, Eric and Kern, Robert and Larson, Eric and Carey, C J and Polat, {\.I}lhan and Feng, Yu and Moore, Eric W. and {VanderPlas}, Jake and Laxalde, Denis and Perktold, Josef and Cimrman, Robert and Henriksen, Ian and Quintero, E. A. and Harris, Charles R. and Archibald, Anne M. and Ribeiro, Ant{\^o}nio H. and Pedregosa, Fabian and {van Mulbregt}, Paul and {SciPy 1.0 Contributors}},
  journal = {Nature Methods},
  title   = {{{SciPy} 1.0: Fundamental Algorithms for Scientific Computing in Python}},
  year    = {2020},
  pages   = {261--272},
  volume  = {17},
  adsurl  = {https://rdcu.be/b08Wh},
  doi     = {10.1038/s41592-019-0686-2},
}

@Misc{BHG2017,
  author = {{Becker \& Hickl GmbH}},
  note   = {v2},
  title  = {{HPM-100-06/07: Ultra-High Speed Hybrid Detectors for TCSPC}},
  year   = {2017},
  url    = {https://www.becker-hickl.com/wp-content/uploads/2018/11/db-hpm-06-07-v02.pdf},
}

@Misc{PhotonscoreGmbH2024,
  author = {{Photonscore GmbH}},
  title  = {{LINTag: High-Throughput Single-Photon-Streaming}},
  year   = {2024},
  url    = {https://www.photonscore.de/wp-content/uploads/2024/09/Photonscore_Datasheet_LINTag.pdf},
}

@InProceedings{Lipinski2011,
  author    = {Lipiński, Maciej and Włostowski, Tomasz and Serrano, Javier and Alvarez, Pablo},
  booktitle = {2011 IEEE International Symposium on Precision Clock Synchronization for Measurement, Control and Communication},
  title     = {White rabbit: a PTP application for robust sub-nanosecond synchronization},
  year      = {2011},
  pages     = {25-30},
  doi       = {10.1109/ISPCS.2011.6070148},
}

@Article{Karl2022,
  author  = {Karl, Sebastian and Zmija, Andreas and Richter, Stefan and Vogel, Naomi and Malyshev, Dmitry and Zink, Adrian and Michel, Thilo and Anton, Gisela and von Zanthier, Joachim and Funk, Stefan},
  journal = {Monthly Notices of the Royal Astronomical Society},
  title   = {Comparing different approaches for stellar intensity interferometry},
  year    = {2022},
  issn    = {0035-8711},
  month   = {02},
  number  = {2},
  pages   = {1722-1729},
  volume  = {512},
  doi     = {10.1093/mnras/stac489},
  eprint  = {https://academic.oup.com/mnras/article-pdf/512/2/1722/42993259/stac489.pdf},
  url     = {https://doi.org/10.1093/mnras/stac489},
}

@Article{Cortina2005,
  author   = {Cortina, Juan},
  journal  = {Astrophysics and Space Science},
  title    = {Status and First Results of the Magic Telescope},
  year     = {2005},
  issn     = {1572-946X},
  month    = {Jun},
  number   = {1},
  pages    = {245-255},
  volume   = {297},
  day      = {01},
  doi      = {10.1007/s10509-005-7627-5},
  url      = {https://doi.org/10.1007/s10509-005-7627-5},
}

@InProceedings{Foffano2021,
  author     = {Foffano, Luca and Carosi, Alessandro and della Volpe, Domenico and Dalchenko, Mykhailo and Heller, Matthieu and Montaruli, Teresa and Abe, Hyuga and Aguasca, Arnau and Agudo, Ivan and Antonelli, Lucio Angelo and Aramo, Carla and Armstrong, Thomas and Artero, Manuel and Asano, Katsuaki and Ashkar, Halim and Aubert, Pierre and Baktash, Ali and Bamba, Aya and Baquero Larriva, Andres and Baroncelli, Leonardo and Barres de Almeida, Ulisses and Barrio, Juan Abel and Batković, Ivana and Becerra Gonzalez, Josefa and Bernardos, Mabel and Berti, Alessio and Biederbeck, Noah and Bigongiari, Ciro and Blanch, Oscar and Bonnoli, Giacomo and Bordas, Pol and Bose, Debanjan and Bulgarelli, Andrea and Burelli, Irene and Buscemi, Mario and Cardillo, Martina and Caroff, Sami and Cassol, Franca and Cerruti, Matteo and Chai, Yating and Cheng, Ks and Chikawa, Michiyuki and Chytka, Ladislav and Contreras, Jose Luis and Cortina, Juan and Costantini, Heide and De Angelis, Alessandro and de Bony de Lavergne, Mathieu and Deleglise, Guillaume and Delgado, Carlos and Delgado Mengual, Jordi and Depaoli, Davide and Di Pierro, Federico and Di Venere, Leonardo and Díaz, Carlos and Dominik, Rune Michael and Dominis Prester, Dijana and Donini, Alice and Dorner, Daniela and Doro, Michele and Elsässer, Dominik and Emery, Gabriel and Escudero, Juan and Fiasson, Armand and Fonseca, Maria Victoria and Freixas Coromina, Lluis and Fukami, Satoshi and Fukazawa, Yasushi and Garcia, Enrique and Garcia López, Ramon and Giglietto, Nicola and Giordano, Francesco and Gliwny, Pawel and Godinovic, Nikola and Green, David and Grespan, Pietro and Gunji, Shuichi and Hackfeld, Jonas and Hadasch, Daniela and Hahn, Alexander and Hassan, Tarek and Hayashi, Kohei and Heckmann, Lea and Herrera Llorente, Javier and Hirotani, Kouichi and Hoffmann, Dirk and Horns, Dieter and Houles, Julien and Hrabovsky, Miroslav and Hrupec, Dario and Hui, David and Hütten, Moritz and Inada, Tomohiro and Inome, Yusuke and Iori, Maurizio and Ishio, Kazuma and Iwamura, Yuki and Jacquemont, Mikael and Jiménez Martínez, Irene and Jouvin, Léa and Jurysek, Jakub and Kagaya, Mika and Karas, Vladimir and Katagiri, Hideaki and Kataoka, Jun and Kerszberg, Daniel and Kobayashi, Yukiho and Kong, Albert and Kubo, Hidetoshi and Kushida, Junko and Lamanna, Giovanni and Lamastra, Alessandra and Le Flour, Thierry and Longo, Francesco and Lopez-Coto, Ruben and López-Moya, Marcos and López-Oramas, Alicia and Luque-Escamilla, Pedro L. and Majumdar, Pratik and MAKARIEV, Martin and Mandat, Dusan and Manganaro, Marina and Mannheim, Karl and Mariotti, Mosè and Marquez, Patricia and Marsella, Giovanni and Martí, Josep and Martinez, Oibar and Martínez, Gustavo and Martinez, Manel and Marusevec, Petra and Mas, Alvaro and Maurin, Gilles and Mazin, Daniel and Mestre Guillen, Enrique and Mićanović, Saša and Miceli, Davide and Miener, Tjark and Miranda, Jose Miguel and Miranda, Luis David Medina and Mirzoyan, Razmik and Mizuno, Tsunefumi and Molina, Edgar and Monteiro, Inocencio and Moralejo, Abelardo and Morcuende, Daniel and Moretti, Elena and Morselli, Aldo and Mrakovcic, Karlo and Murase, Kohta and Nagai, Andrii and Nakamori, Takeshi and Nickel, Lukas and Nieto, Daniel and Nievas, Mireia and Nishijima, Kyoshi and Noda, Koji and Nosek, Dalibor and Nöthe, Maximilian and Nozaki, Seiya and Ohishi, Michiko and Ohtani, Yoshiki and Oka, Tomohiko and Okazaki, Nao and Okumura, Akira and Orito, Reiko and Otero-Santos, Jorge and Palatiello, Michele and Paneque, David and Paoletti, Riccardo and Paredes, Josep Maria and Pavletić, Lovro and Pech, Miroslav and Pecimotika, Mario and Poireau, Vincent and Polo, Miguel and Prandini, Elisa and Prast, Julie and Priyadarshi, Chaitanya and Prouza, Michael and Rando, Riccardo and Rhode, Wolgang and Ribó, Marc and Rizi, Vincenzo and Rugliancich, Andrea and Ruiz, Jose Enrique and Saito, Takayuki and Sakurai, Shunsuke and Sanchez, David and Šarić, Toni and Saturni, Francesco Gabriele and Scherpenberg, Juliane and Schleicher, Bernd and Schubert, Jan Lukas and Schussler, Fabian and Schweizer, Thomas and Seglar Arroyo, Monica and Shellard, Ronald Cintra and Sitarek, Julian and Sliusar, Vitalii and Spolon, Alessia and Strišković, Jelena and Strzys, Marcel C. and Suda, Yusuke and Sunada, Yuji and Tajima, Hiroyasu and Takahashi, Mitsunari and Takahashi, Hiromitsu and Takata, Jumpei and Takeishi, Ryuji and Tam, P. Thomas and Tanaka, Shuta and Tateishi, Dai and Tejedor, Luis Ángel and Temnikov, Petar and Terada, Yukikatsu and Terzic, Tomislav and Teshima, Masahiro and Tluczykont, Martin and Tokanai, Fuyuki and Torres, Diego F. and Travnicek, Petr and Truzzi, Stefano and Vacula, Martin and VAZQUEZ ACOSTA, Monica and VERGUILOV, Vassil and Verna, Gaia and Viale, Ilaria and Vigorito, Carlo Francesco and Vitale, Vincenzo and Vovk, Ievgen and Vuillaume, Thomas and Walter, Roland and Will, Martin and Yamamoto, Tokonatsu and Yamazaki, Ryo and Yoshida, Tatsuo and Yoshikoshi, Takanori and Zarić, Darko},
  booktitle  = {Proceedings of 37th International Cosmic Ray Conference — PoS(ICRC2021)},
  title      = {Monitoring the pointing of the Large Size Telescope prototype using star reconstruction in the Cherenkov camera},
  year       = {2021},
  month      = jul,
  pages      = {712},
  publisher  = {Sissa Medialab},
  series     = {ICRC2021},
  collection = {ICRC2021},
  doi        = {10.22323/1.395.0712},
  url        = {http://dx.doi.org/10.22323/1.395.0712},
}

@Misc{Dalchenko2023,
  author        = {Mykhailo Dalchenko and Matthieu Heller},
  title         = {Imaging Atmospheric Cherenkov Telescopes pointing determination using the trajectories of the stars in the field of view},
  year          = {2023},
  archiveprefix = {arXiv},
  eprint        = {2310.04101},
  primaryclass  = {astro-ph.IM},
  url           = {https://arxiv.org/abs/2310.04101},
}

@TechReport{SED2023,
  author      = {{Safran Electronics \& Defense}},
  institution = {Safran Electronics \& Defense},
  title       = {White Rabbit LEN: The cost effective node to distribute synchronization},
  year        = {2023},
  month       = {February},
  note        = {Accessed: 2026-01-08},
  type        = {Datasheet},
  url         = {https://safran-navigation-timing.com/wp-content/uploads/2023/03/White-Rabbit-LEN-SAFRAN-Datasheet.pdf},
}

@Article{Guerin2017,
  author   = {Guerin, W. and Dussaux, A. and Fouché, M. and Labeyrie, G. and Rivet, J.-P. and Vernet, D. and Vakili, F. and Kaiser, R.},
  journal  = {Monthly Notices of the Royal Astronomical Society},
  title    = {{Temporal intensity interferometry: photon bunching in three bright stars}},
  year     = {2017},
  issn     = {0035-8711},
  month    = {10},
  number   = {4},
  pages    = {4126-4132},
  volume   = {472},
  doi      = {10.1093/mnras/stx2143},
  eprint   = {https://academic.oup.com/mnras/article-pdf/472/4/4126/21072345/stx2143.pdf},
  url      = {https://doi.org/10.1093/mnras/stx2143},
}

@Article{Zampieri2021,
  author   = {Zampieri, Luca and Naletto, Giampiero and Burtovoi, Aleksandr and Fiori, Michele and Barbieri, Cesare},
  journal  = {Monthly Notices of the Royal Astronomical Society},
  title    = {{Stellar intensity interferometry of Vega in photon counting mode}},
  year     = {2021},
  issn     = {0035-8711},
  month    = {05},
  number   = {2},
  pages    = {1585-1594},
  volume   = {506},
  doi      = {10.1093/mnras/stab1387},
  eprint   = {https://academic.oup.com/mnras/article-pdf/506/2/1585/38968815/stab1387.pdf},
  url      = {https://doi.org/10.1093/mnras/stab1387},
}

@Article{Horch2022,
  author    = {Horch, Elliott P. and Weiss, Samuel A. and Klaucke, Paul M. and Pellegrino, Richard A. and Rupert, Justin D.},
  journal   = {The Astronomical Journal},
  title     = {Observations with the Southern Connecticut Stellar Interferometer. I. Instrument Description and First Results},
  year      = {2022},
  month     = {jan},
  number    = {2},
  pages     = {92},
  volume    = {163},
  doi       = {10.3847/1538-3881/ac43bb},
  publisher = {The American Astronomical Society},
  url       = {https://doi.org/10.3847/1538-3881/ac43bb},
}

@Article{Mozdzen2025,
  author   = {Mozdzen, Thomas J and Scott, Richard M and Mauskopf, Philip D and Rodriguez, Ricardo R},
  journal  = {Monthly Notices of the Royal Astronomical Society},
  title    = {Intensity interferometer results on Sirius with 0.25 m telescopes},
  year     = {2025},
  issn     = {0035-8711},
  month    = {02},
  number   = {3},
  pages    = {2527-2534},
  volume   = {537},
  doi      = {10.1093/mnras/staf226},
  eprint   = {https://academic.oup.com/mnras/article-pdf/537/3/2527/61779152/staf226.pdf},
  url      = {https://doi.org/10.1093/mnras/staf226},
}

@Book{Macleod2010,
  author    = {Macleod, H Angus and Macleod, H Angus},
  publisher = {CRC press},
  title     = {Thin-film optical filters},
  year      = {2010},
}

@Article{Guerin2025,
  author  = {Guerin, William and Hugbart, Mathilde and Tolila, Sarah and Matthews, Nolan and Lai, Olivier and Rivet, Jean-Pierre and Labeyrie, Guillaume and Kaiser, Robin},
  journal = {Comptes Rendus. Physique},
  title   = {Stellar intensity interferometry in the photon-counting regime},
  year    = {2025},
  number  = {G1},
  pages   = {659--679},
  volume  = {26},
}

@Article{Dalal2024,
  author    = {Dalal, Neal and Galanis, Marios and Gammie, Charles and Gralla, Samuel E. and Murray, Norman},
  journal   = {Phys. Rev. D},
  title     = {Probing ${H}_{0}$ and resolving AGN disks with ultrafast photon counters},
  year      = {2024},
  month     = {Jun},
  pages     = {123029},
  volume    = {109},
  doi       = {10.1103/PhysRevD.109.123029},
  issue     = {12},
  numpages  = {21},
  publisher = {American Physical Society},
  url       = {https://link.aps.org/doi/10.1103/PhysRevD.109.123029},
}

@Article{Cortina2025,
  author    = {Cortina, J. and Cifuentes Santos, A. and Hassan, T. and Frías García-Lago, F.},
  journal   = {Publications of the Astronomical Society of the Pacific},
  title     = {A Butterfly’s Eye Camera for Intensity Interferometry with Cherenkov Telescopes},
  year      = {2025},
  month     = {dec},
  number    = {12},
  pages     = {125004},
  volume    = {137},
  doi       = {10.1088/1538-3873/ae20c0},
  publisher = {The Astronomical Society of the Pacific},
  url       = {https://doi.org/10.1088/1538-3873/ae20c0},
}

@Article{Loesler2013,
  author   = {L{\"o}sler, Michael and Haas, R{\"u}diger and Eschelbach, Cornelia},
  journal  = {Journal of Geodesy},
  title    = {Automated and continual determination of radio telescope reference points with sub-mm accuracy: results from a campaign at the Onsala Space Observatory},
  year     = {2013},
  issn     = {1432-1394},
  month    = {Aug},
  number   = {8},
  pages    = {791-804},
  volume   = {87},
  day      = {01},
  doi      = {10.1007/s00190-013-0647-y},
  url      = {https://doi.org/10.1007/s00190-013-0647-y},
}

@ARTICLE{PHOENIX_2025,
       author = {{Hauschildt}, P.~H. and {Barman}, T. and {Baron}, E. and {Aufdenberg}, J.~P. and {Schweitzer}, A.},
        title = "{The NewEra model grid}",
      journal = {\aap},
         year = 2025,
        month = jun,
       volume = {698},
          eid = {A47},
        pages = {A47},
          doi = {10.1051/0004-6361/202554171},
archivePrefix = {arXiv},
       eprint = {2504.17597},
 primaryClass = {astro-ph.SR},
       adsurl = {https://ui.adsabs.harvard.edu/abs/2025A&A...698A..47H}
      }

@ARTICLE{Monnier_2012,
       author = {{Monnier}, J.~D. and {Che}, Xiao and {Zhao}, Ming and {Ekstr{\"o}m}, S. and {Maestro}, V. and {Aufdenberg}, Jason and {Baron}, F. and {Georgy}, C. and {Kraus}, S. and {McAlister}, H. and {Pedretti}, E. and {Ridgway}, S. and {Sturmann}, J. and {Sturmann}, L. and {ten Brummelaar}, T. and {Thureau}, N. and {Turner}, N. and {Tuthill}, P.~G.},
        title = "{Resolving Vega and the Inclination Controversy with CHARA/MIRC}",
      journal = {\apjl},
         year = 2012,
        month = dec,
       volume = {761},
       number = {1},
          eid = {L3},
        pages = {L3},
          doi = {10.1088/2041-8205/761/1/L3},
archivePrefix = {arXiv},
       eprint = {1211.6055},
 primaryClass = {astro-ph.SR},
       adsurl = {https://ui.adsabs.harvard.edu/abs/2012ApJ...761L...3M}
}

@ARTICLE{Vega_2006A,
       author = {{Aufdenberg}, J.~P. and {M{\'e}rand}, A. and {Coud{\'e} du Foresto}, V. and {Absil}, O. and {Di Folco}, E. and {Kervella}, P. and {Ridgway}, S.~T. and {Berger}, D.~H. and {ten Brummelaar}, T.~A. and {McAlister}, H.~A. and {Sturmann}, J. and {Sturmann}, L. and {Turner}, N.~H.},
        title = "{First Results from the CHARA Array. VII. Long-Baseline Interferometric Measurements of Vega Consistent with a Pole-On, Rapidly Rotating Star}",
      journal = {\apj},
         year = 2006,
        month = jul,
       volume = {645},
       number = {1},
        pages = {664-675},
          doi = {10.1086/504149},
archivePrefix = {arXiv},
       eprint = {astro-ph/0603327},
 primaryClass = {astro-ph},
       adsurl = {https://ui.adsabs.harvard.edu/abs/2006ApJ...645..664A}
}

@ARTICLE{hipparcos,
       author = {{van Leeuwen}, F.},
        title = "{Validation of the new Hipparcos reduction}",
      journal = {\aap},
         year = 2007,
        month = nov,
       volume = {474},
       number = {2},
        pages = {653-664},
          doi = {10.1051/0004-6361:20078357},
archivePrefix = {arXiv},
       eprint = {0708.1752},
 primaryClass = {astro-ph},
       adsurl = {https://ui.adsabs.harvard.edu/abs/2007A&A...474..653V}
}
\begin{appendix}





\end{appendix}
\end{document}